\input phyzzx
\overfullrule=0pt
\def\sqr#1#2{{\vcenter{\hrule height.#2pt
      \hbox{\vrule width.#2pt height#1pt \kern#1pt
          \vrule width.#2pt}
      \hrule height.#2pt}}}

\def\square{{\mathchoice{\sqr84}{\sqr84}{\sqr{5.0}3}{\sqr{3.5}3}}}
\def\dA{\mathop\square\nolimits}
\def\Lambdato{{\buildrel{\Lambda\rightarrow\infty}\over=}}
%
\REF\JAC{%
I. Jack and D. R. T. Jones,
Liverpool University Preprint LTH-400, hep-ph/9707278.}
\REF\DEW{%
B. S. DeWitt,
{\sl Dynamical Theory of Group and Fields\/} (Gordon and Breach, New
York, 1978);
Phys.\ Rev.\ {\bf 162} (1967), 1195.\nextline
J. Honerkamp,
Nucl.\ Phys.\ {\bf B36} (1971), 130; Nucl.\ Phys.\ {\bf B48} (1972),
269.\nextline
G. 't Hooft,
Nucl.\ Phys.\ {\bf B62} (1973), 444.}
\REF\ABB{%
L. F. Abbott,
Nucl.\ Phys.\ {\bf B185} (1981), 189.\nextline
L. F. Abbott, M. T. Grisaru and R. K. Schaefer,
Nucl.\ Phys.\ {\bf B229} (1983), 372.}
\REF\ICH{%
S. Ichinose and M. Omote,
Nucl.\ Phys.\ {\bf B203} (1982), 221.}
\REF\SAL{%
A. Salam and J. Strathdee,
Phys.\ Rev.\ {\bf D11} (1975), 1521.\nextline
S. Ferrara and O. Piguet,
Nucl.\ Phys.\ {\bf B93} (1975), 261.}
\REF\GRI{%
M. T. Grisaru, W. Siegel and M. Ro\v cek,
Nucl.\ Phys.\ {\bf B159} (1979), 429.}
\REF\SIE{%
W. Siegel, Phys.\ Lett.\ {\bf 84B} (1979), 193; Phys.\ Lett.\
{\bf 94B} (1980),
37.
\nextline
D. M. Capper, D. R. T. Jones and P. van Nieuwenhuizen,
Nucl.\ Phys.\ {\bf B167} (1980), 479.\nextline
L. V. Avdeev, G. V. Ghochia and A. A. Vladiminov,
Phys.\ Lett.\ {\bf 105B} (1981), 272.\nextline
L. V. Avdeev and A. A. Vladiminov,
Nucl.\ Phys.\ {\bf B219} (1983), 262.}
\REF\WES{%
P. West,
Nucl.\ Phys.\ {\bf B268} (1986), 113, and references therein.}
\REF\GAI{%
M. K. Gaillard,
Phys.\ Lett.\ {\bf B342} (1995), 125; Phys.\ Lett.\ {\bf B347}
(1995), 284.}
\REF\FRO{%
S. A. Frolov and A. A. Slavnov,
Phys.\ Lett.\ {\bf B309} (1993), 344.\nextline
R. Narayanan and H. Neuberger,
Phys.\ Lett.\ {\bf B302} (1993), 62.\nextline
S. Aoki and Y. Kikukawa,
Mod.\ Phys.\ Lett.\ {\bf A8} (1993), 3517.\nextline
K. Fujikawa,
Indian J.\ Phys.\ {\bf 70A} (1996), 275.\nextline
K. Okuyama and H. Suzuki,
Phys.\ Lett.\ {\bf B382} (1996), 117; Prog.\ Theor.\ Phys. {\bf 98}
(1997), 463.
\nextline
L. N. Chang and C. Soo,
Phys.\ Rev.\ {\bf D55} (1997), 2410.}
\REF\BAR{%
W. A. Bardeen,
Phys.\ Rev.\ {\bf 184} (1969), 1848.}
\REF\FUJ{%
K. Fujikawa,
Phys.\ Rev.\ {\bf D29} (1984), 285; Phys.\ Rev.\ {\bf D48} (1993),
3922; Nucl.\ Phys.\ {\bf B428} (1994), 169.}
\REF\BAN{%
H. Banerjee, R. Banerjee and P. Mitra,
Z. Phys.\ {\bf C32} (1986), 445.}
\REF\BARD{%
L. Alvarez-Gaum\'e and E. Witten,
Nucl.\ Phys.\ {\bf B234} (1983), 269.\nextline
W. A. Bardeen and B. Zumino,
Nucl.\ Phys.\ {\bf B244} (1984), 421.}
\REF\KON{%
K. Konishi and K. Shizuya,
Nuovo Cim.\ {\bf 90A} (1985), 111.}
\REF\CLA{%
T. E. Clark, O. Piguet and K. Sibold,
Nucl.\ Phys.\ {\bf B159} (1979), 1.}
\REF\GAT{%
S. J. Gates, M. T. Grisaru, M. Ro\v cek and W. Siegel,
{\sl Superspace or One Thousand-and-One Lessons in Supersymmetry\/}
(Benjamin/Cummings, Reading, Mass., 1983).}
\REF\FER{%
S. Ferrara and B. Zumino,
Nucl.\ Phys.\ {\bf B87} (1975), 207.}
\REF\CLAR{%
T. E. Clark, O. Piguet and K. Sibold,
Nucl.\ Phys.\ {\bf B143} (1978), 445; Nucl.\ Phys.\ {\bf B172}
(1980), 201.}
\REF\LAN{%
W. Lang,
Nucl.\ Phys.\ {\bf B150} (1979), 201.}
\REF\MAR{%
S. Marculescu and L. Mezincescu,
Nucl.\ Phys.\ {\bf B181} (1981), 127.}
\REF\SHI{%
K. Shizuya,
Phys.\ Rev.\ {\bf D35} (1987), 1848.}
\REF\KONI{%
K. Konishi,
Phys.\ Lett.\ {\bf 135B} (1984), 439.}
\REF\ITO{%
H. Itoyama and R. Razzaghe-Ashrafi,
Phys.\ Lett.\ {\bf B291} (1992), 268.}
\REF\DVA{%
G. Dvali and M. Shifman,
Phys.\ Lett.\ {\bf B396} (1997), 64.\nextline
A. Kovner, M. Shifman and A. Smilga,
Phys.\ Rev.\ {\bf D56} (1997), 7978.\nextline
B. Chibisov and M. Shifman,
Phys.\ Rev.\ {\bf D56} (1997), 7990.}
\REF\WESS{%
J. Wess and J. Bagger,
{\sl Supersymmetry and Supergravity\/} (Princeton University Press,
Princeton, 1992).}
\REF\GRIS{%
M. T. Grisaru and W. Siegel,
Nucl.\ Phys.\ {\bf B201} (1979), 292.}
\REF\WEST{%
See, for example, P. West,
{\sl Introdunction to Supersymmetry and Supergravity\/} (World
Scientific, Singapore, 1990).}
\REF\NIE{%
N. K. Nielsen,
Nucl.\ Phys.\ {\bf B140} (1978), 499.\nextline
R. E. Kallosh,
Nucl.\ Phys.\ {\bf B141} (1978), 141.}
\REF\SHIZ{%
K. Shizuya and Y. Yasui,
Phys.\ Rev.\ {\bf D29} (1984), 1160.}
\REF\SCH{%
J. Schwinger,
Phys.\ Rev.\ {\bf 82} (1951), 664.}
\REF\NIEL{%
N. K. Nielsen,
Nucl.\ Phys.\ {\bf B244} (1984), 499.\nextline
E. Guadagnini, K. Konishi and M. Mintchev,
Phys.\ Lett.\ {\bf 157B} (1985), 37.\nextline
K. Harada and K. Shizuya,
Phys.\ Lett.\ {\bf 162B} (1985), 322.\nextline
R. Garreis, M. Scholl and J. Wess,
Z. Phys.\ {\bf C28} (1985), 628.}
\REF\PIG{%
O. Piguet and K. Sibold,
Nucl.\ Phys.\ {\bf B247} (1984), 484.\nextline
T. E. Clark and S. Love,
Phys.\ Lett.\ {\bf 138B} (1984), 289.\nextline
D. Nemeschansky and R. Rohm,
Nucl.\ Phys.\ {\bf B249} (1985), 157.\nextline
G. Girardi, R. Grimm and R. Stora,
Phys.\ Lett.\ {\bf 156B} (1985), 203.}
\REF\HAY{%
T. Hayashi, Y. Ohshima, K. Okuyama and H. Suzuki, Ibaraki University
Preprint IU-MSTP/29; hep-th/9805068.}
\REF\FAY{%
P. Fayet and J. Iliopoulos,
Phys.\ Lett.\ {\bf 51B} (1974), 461.\nextline
W. Fisher, N. P. Nilles, J. Polchinski, S. Raby and L. Susskind,
Phys.\ Rev.\ Lett.\ {\bf 47} (1981), 757.}
\REF\THO{%
G. 't Hooft,
Phys.\ Rev.\ Lett.\ {\bf 37} (1976), 8.}
\REF\SOH{%
M. Sohnius and P. West,
Phys.\ Lett.\ {\bf 105B} (1981), 353.}
\REF\FUJI{%
K. Fujikawa and K. Okuyama,
Nucl.\ Phys.\ {\bf B521} (1998), 401.}
\REF\GRISA{%
M. Grisaru and P. West,
Nucl.\ Phys.\ {\bf B254} (1985), 249.}
\REF\SHIF{%
M. Shifman and A. Vainshtein,
Nucl.\ Phys.\ {\bf B277} (1986), 456.}
%
\pubnum={IU-MSTP/27; hep-th/9801062}
\date={January 1998}
\titlepage
\title{Invariant Regularization of Supersymmetric Chiral Gauge Theory}
\author{%
Takuya Hayashi,\foot{E-mail: hayashi@physun1.sci.ibaraki.ac.jp}
Yoshihisa Ohshima\foot{E-mail: ohshima@mito.ipc.ibaraki.ac.jp}\break
Kiyoshi Okuyama\foot{E-mail: okuyama@mito.ipc.ibaraki.ac.jp}
and Hiroshi Suzuki\foot{E-mail: hsuzuki@mito.ipc.ibaraki.ac.jp}}
\address{%
Department of Physics, Ibaraki University, Mito 310-0056, Japan}
\abstract{%
We formulate a manifestly supersymmetric gauge covariant
regularization of supersymmetric chiral gauge theories. In our scheme,
the effective action in the superfield background field method
{\it above\/} one-loop is always supersymmetric and gauge invariant.
The gauge anomaly has a covariant form and can emerge only in
one-loop diagrams with all the external lines being the background
gauge superfield. We also present several illustrative applications
in the one-loop approximation: the self-energy part of the chiral
multiplet and of the gauge multiplet; the super-chiral anomaly and the
superconformal anomaly; as the corresponding anomalous commutators,
the Konishi anomaly and an anomalous supersymmetric transformation law
of the supercurrent (the ``central extension'' of $N=1$~supersymmetry
algebra) and of the $R$-current.
\endpage
\chapter{Introduction}
Obtaining a regularization scheme which is invariant under preferred
symmetries (such as the gauge symmetry) is important. As a matter of
principle, the existence of such a regularization directly shows that
these symmetries have no intrinsic quantum anomaly. Even at the
practical level, use of such a regularization (at least conceptually)
simplifies various calculations, because it avoids the introduction
of non-invariant counter-terms necessary to recover the
Ward-Takahashi identity. Such a regularization also automatically
leads to correct physical predictions in view of the preferred or
imposed symmetries; a well-known example is the anomalous divergence
of the axial current (the Adler-Bell-Jackiw anomaly) which would not
appear without imposing the gauge invariance.

In this paper, we formulate an invariant regularization of
supersymmetric {\it chiral\/} gauge theories. As a recent review of
the issues relating to regularization of supersymmetric theories,
see Ref.~[\JAC] and references cited therein. For definiteness, we
set up a regularization scheme for the effective action in the
background field method~[\DEW--\ICH]. Once this is done, $S$-matrix
elements can also be constructed~[\ABB]. Although our scheme is
perturbative in nature, it possesses the following properties. (1)~It
manifestly preserves supersymmetry at every step, being formulated in
terms of the superfield~[\SAL,\GRI] in an exactly four-dimensional
spacetime. In particular, there is no ambiguity associated
with~$\gamma_5$ and the totally anti-symmetric tensor, unlike the
dimensional reduction~[\SIE]. In this respect, our scheme is similar
to the supersymmetric higher covariant derivative
regularization~[\WES]. However our scheme regularizes one-loop
diagrams as well. (2)~It is manifestly gauge covariant, unlike the
conventional Pauli-Villars regularization (for the Pauli-Villars
regularization in supersymmetric theories, see Ref.~[\GAI]); it is
related to the generalized Pauli-Villars regularization~[\FRO]. Here,
by ``gauge covariance,'' we mean covariance under the background gauge
transformation~[\GRI] and not under the BRST transformation. In fact,
in our scheme, the effective action {\it above\/} one-loop is always
gauge invariant, and a possible breaking of the gauge symmetry due to
the gauge anomaly~[\BAR] can emerge only in one-loop diagrams in which
all the external lines are the background gauge superfield. The
one-loop diagrams on the other hand are regularized gauge
covariantly~[\FUJ,\BAN]. Hence the gauge anomaly has a covariant
form~[\FUJ--\KON]. When the gauge representation is free of the gauge
anomaly, however, the gauge invariance is restored and our scheme as
it stands provides a gauge invariant regularization.

The organization of this paper is as follows. In \S2, the superfield
background field method is summarized in a somewhat different
representation from~[\GRI]. This representation is more suitable for
our purpose. On the basis of this method, we diagonalize a part of the
action which is quadratic in quantum fields in the presence of the
background gauge field. This gives us a ``partially diagonalized''
propagator of quantum fields (the first half of \S3). In the second
half of \S3, our regularization scheme is formulated by using the
propagators thus obtained. It is explained how the supersymmetry and
the background gauge invariance or covariance are respected in the
process of regularization. Sections 4 and~5 are devoted to
illustrative applications in the one-loop approximation. In \S4, we
present a somewhat detailed evaluation of the two-point one particle
irreducible (1PI) functions, the self-energy part of the chiral
multiplet and of the gauge multiplet (the vacuum polarization
tensor). In \S5, the super-chiral anomaly~[\KON--\GAT] and the
superconformal anomaly~[\FER--\SHI] are evaluated. Since our scheme
also provides a supersymmetric gauge covariant definition of
composite operators, an evaluation of anomalies in the form of an
anomalous supersymmetric transformation law~[\KONI--\DVA] as well as
in the form of current non-conservation is straightforward and
transparent. Section~6 is devoted to a conclusion. Details concerning
the evaluation of anomalous factors are summarized in the Appendix.

Our convention is basically that of Ref.~[\WESS], unless otherwise
stated. In particular, the signature of the metric
is~$({-}{+}{+}{+})$. For simplicity of presentation, we assume the
gauge representation~$R$ of the chiral multiplet, which will be
denoted as~$T^a$, is irreducible. The normalization of the gauge
generator is $[T^a,T^b]=it^{abc}T^c$, $\tr T^aT^b=T(R)\delta^{ab}$,
$(T^a)_{ij}(T^a)_{jk}=C(R)\delta_{ik}$ and
$t^{acd}t^{bcd}=C_2(G)\delta^{ab}$.

\chapter{Superfield background field method}
We set up our scheme on the basis of the background field
method~[\DEW--\ICH]. The reason is that the background field
method allows us to treat the gauge field and the matter field on an
equal footing in view of the gauge covariance. Moreover, to make the
supersymmetry manifest, we utilize the superfield background field
method. Its basic framework is presented in Ref.~[\GRI] (see
also Ref.~[\GRIS]). However, in view of the ``full-chiral''
representation~[\WESS] that is commonly used, it is rather convenient
to work directly with the following ``quantum-chiral
background-chiral'' representation. This representation differs from
the quantum-chiral background-vector representation adapted in
Ref.~[\GRI]. Therefore, it seems helpful to summarize this
representation while explaining our original notation.

We consider the most general renormalizable supersymmetric model,
whose classical action is given by~[\WESS]\foot{%
The generic coordinate of the superspace is denoted as
$z=(x^m,\theta^\alpha,\overline\theta_{\dot\alpha})$. The full
superspace integration measure and the chiral superspace measure are
abbreviated as $d^8z=d^4xd^2\theta d^2\overline\theta$ and
$d^6z=d^4xd^2\theta$, respectively.}
$$
   S={1\over2T(R)}\int d^6z\,\tr W^\alpha W_\alpha
   +\int d^8z\,\Phi^\dagger e^V\Phi
   +\int d^6z\,\left({1\over2}\Phi^T m\Phi+{1\over3}g\Phi^3\right)
   +{\rm h.c.}
\eqn\twoxone
$$
In our quantum-chiral background-chiral representation,\foot{%
The quantum fields $V_Q$~in this representation and $V^Q$~in the
quantum-chiral background-vector representation~[\GRI] are related as
$V_Q=e^{-g\overline W^B}gV^Qe^{g\overline W^B}$, where $W^B$~is the
background field in Ref.~[\GRI]. Our background field~$V_B$ is given
by $e^{V_B}=e^{gW^B}e^{g\overline W^B}$.}
the vector and the chiral superfields are split as
$$
   e^V=e^{V_B}e^{V_Q},\qquad\Phi=\Phi_B+\Phi_Q.
\eqn\twoxtwo
$$
Here, the subscripts $B$ and~$Q$ represent the background field and
the quantum field, respectively. We shall regard $V_B$~as a vector
superfield, and thus $V_Q$~is {\it not\/} a vector superfield.
Instead, its conjugate is given by
$$
   V_Q^\dagger=e^{V_B}V_Q e^{-V_B}.
\eqn\twoxthree
$$
In this quantum-chiral background-chiral representation, the original
gauge transformation~[\WESS]
$$
   e^{V'}=e^{-i\Lambda^\dagger}e^Ve^{i\Lambda},
   \qquad\Phi'=e^{-i\Lambda}\Phi,
\eqn\twoxfour
$$
where $\Lambda=T^a\Lambda^a$ is a chiral superfield
$\overline D_{\dot\alpha}\Lambda=0$, is realized in the following two
different ways. (i) By the quantum field transformation:
$$
   V_B'=V_B,
   \qquad e^{V_Q'}
   =\left(e^{-V_B}e^{-i\Lambda^\dagger}e^{V_B}\right)e^{V_Q}
   e^{i\Lambda},
   \quad\Phi'=e^{-i\Lambda}\Phi.
\eqn\twoxfive
$$
(ii) By the background field transformation:
$$
   e^{V_B'}=e^{-i\Lambda^\dagger}e^{V_B}e^{i\Lambda},
   \qquad V_Q'=e^{-i\Lambda}V_Qe^{i\Lambda},
   \quad\Phi'=e^{-i\Lambda}\Phi.
\eqn\twoxsix
$$
In both transformations, the gauge parameter~$\Lambda$ is simply a
{\it chiral\/} superfield, whence the name of the representation.

Next, we introduce the background covariant spinor derivative symbol:
$$
   \nabla_\alpha\equiv e^{-V_B}D_\alpha e^{V_B}\quad{\rm and}
   \quad\overline D_{\dot\alpha}.
\eqn\twoxseven
$$
Since the gauge parameter~$\Lambda$ in~\twoxsix\ is chiral, both of
these operators transform as
$\nabla'=e^{-i\Lambda}\nabla e^{i\Lambda}$ under the background field
transformation. The vector covariant derivative symbol is also defined
by the anti-commutator:
$$
   \{\nabla_\alpha,\overline D_{\dot\alpha}\}
   \equiv-2i\sigma^m_{\alpha\dot\alpha}\nabla_m.
\eqn\twoxeight
$$
Then, on the gauge representation~$R$, the covariant derivative is
defined by
$$
   {\cal D}_\alpha\Phi\equiv\nabla_\alpha\Phi,
   \qquad{\cal D}_m\Phi\equiv\nabla_m\Phi,
\eqn\twoxnine
$$
where $\Phi$~is a generic field in the representation~$R$. Clearly
these operations have a covariant meaning under the background field
transformation~\twoxsix. On the other hand, the quantum field~$V_Q$
transforms as the adjoint representation under the background
transformation~\twoxsix. In such a representation, the covariant
derivative is defined by
$$
   {\cal D}_\alpha V\equiv[\nabla_\alpha,V\},
   \qquad{\cal D}_mV\equiv[\nabla_m,V],
\eqn\twoxten
$$
where a(n) (anti-)commutator is used when $V$~is
Grassmann-even(-odd). It is again clear that \twoxten~has a
background gauge covariant meaning.

Expressions become even simpler with use of the adjoint gauge
representation matrix, which is defined by
$$
   ({\cal T}^a)^{bc}\equiv-it^{abc},
   \qquad\tr{\cal T}^a{\cal T}^b=C_2(G)\delta^{ab}.
\eqn\twoxeleven
$$
With this convention, the covariant derivative in the adjoint
representation~\twoxten\ can be written as
$$
   {\cal D}_\alpha V=T^a(\widetilde\nabla_\alpha V)^a,
   \qquad{\cal D}_m V=T^a(\widetilde\nabla_m V)^a,
\eqn\twoxtwelve
$$
where a component of the covariant derivative is defined by
$$
   (\widetilde\nabla_\alpha V)^a
   \equiv(e^{-{\cal V}_B})^{ab}D_\alpha(e^{{\cal V}_B})^{bc}V^c,
   \qquad
   \{\widetilde\nabla_\alpha,\overline D_{\dot\alpha}\}
   =-2i\sigma^m_{\alpha\dot\alpha}\widetilde\nabla_m,
\eqn\twoxthirteen
$$
and ${\cal V}_B$~is the background gauge superfield in the adjoint
representation:
$$
   {\cal V}_B\equiv{\cal T}^aV_B^a.
\eqn\twoxfourteen
$$
The similarity of~\twoxthirteen\ and the covariant derivative
\twoxseven\ and~\twoxnine\ is obvious.

Now the essence of the background field method~[\DEW,\ABB,\ICH] is to
use the gauge fixing condition which is covariant under the
background gauge transformation~\twoxsix. Therefore, as usual, we
impose the Lorentz-type gauge fixing condition and its conjugate:
$$
   \overline D^2V_Q=f,
   \qquad{\cal D}^2V_Q=e^{-V_B}f^\dagger e^{V_B}.
\eqn\twoxfifteen
$$
Note that the gauge fixing function $f$~is a chiral superfield:
$\overline D_{\dot\alpha}f=0$. Then the standard procedure~[\WEST]
gives rise to the gauge fixing term and the ghost-anti-ghost term:
$$
\eqalign{
   S'&=-{\xi\over8T(R)}\int d^8z\,\tr(\overline D^2V_Q)({\cal D}^2V_Q)
\cr
   &\quad+{1\over T(R)}\int d^8z\,
   \tr(e^{-V_B}c^{\prime\dagger}e^{V_B}+c^\prime)
\cr
   &\qquad\qquad\qquad\times
   {\cal L}_{V_Q/2}\cdot
   \left[(c+e^{-V_B}c^\dagger e^{V_B})
    +\coth({\cal L}_{V_Q/2})\cdot
   (c-e^{-V_B}c^\dagger e^{V_B})\right]
\cr
   &\quad+{1\over T(R)}\int d^8z\,\tr e^{-V_B}b^\dagger e^{V_B}b,
\cr
}
\eqn\twoxsixteen
$$
where $\xi$~is the gauge parameter. By construction, this action is
invariant under the background field transformation, \twoxsix\ and
$b'=e^{-i\Lambda}be^{i\Lambda}$ etc. Note that, since the
parameter~$\Lambda$ of the quantum field transformation~\twoxfive\ is
chiral and the gauge fixing function~$f$ in~\twoxfifteen\ is also
chiral, all the ghost~$c$, anti-ghost~$c^\prime$, and Nielsen-Kallosh
ghost~$b$~[\NIE] are simply chiral superfields:
$\overline D_{\dot\alpha}c=\overline D_{\dot\alpha}c^\prime=
\overline D_{\dot\alpha}b=0$.

To carry out perturbative calculations, we expand the total
action~$S_T\equiv S+S'$ in powers of the quantum fields as
$S_T=S_{T0}+S_{T1}+S_{T2}+S_{T3}+\cdots$. The zeroth order action has
the same form as the classical action~$S$,
$$
\eqalign{
   S_{T0}&={1\over2T(R)}\int d^6z\,\tr W_B^\alpha W_{B\alpha}
   +\int d^8z\,\Phi_B^\dagger e^{V_B}\Phi_B
\cr
   &\quad+\int d^6z\,
   \left({1\over2}\Phi_B^Tm\Phi_B+{1\over3}g\Phi_B^3\right)
   +{\rm h.c.},
\cr
}
\eqn\twoxseventeen
$$
where the background field strength~$W_{B\alpha}$ has been defined by
$$
   W_{B\alpha}\equiv-{1\over4}\overline D^2
   (e^{-V_B}D_\alpha e^{V_B})
   =-{1\over4}[\overline D_{\dot\alpha},
                   \{\overline D^{\dot\alpha},\nabla_\alpha\}].
\eqn\twoxeighteen
$$
The last expression is convenient for various calculations, and it
shows that $W_{B\alpha}$~is in fact a gauge covariant object. The
first order action~$S_{T1}$ is given by
$$
\eqalign{
   S_{T1}&=-{1\over T(R)}\int d^8z\,\tr V_Q{\cal D}^\alpha W_{B\alpha}
\cr
   &\quad+\int d^8z\,(\Phi_B^\dagger e^{V_B}V_Q\Phi_B
                +\Phi_Q^\dagger e^{V_B}\Phi_B
                +\Phi_B^\dagger e^{V_B}\Phi_Q)
\cr
   &\quad+\int d^6z\,(\Phi_Q^Tm\Phi_B+g\Phi_Q\Phi_B^2)
   +{\rm h.c.}
\cr
}
\eqn\twoxnineteen
$$
Since the first order action does not contribute to 1PI diagrams,
$S_{T1}$~can be neglected in the following discussion.

Let us study the action quadratic in the quantum fields, $S_{T2}$. We
decompose it as $S_{T2}=S_{T2}^{\rm gauge}+S_{T2}^{\rm chiral}
+S_{T2}^{\rm mix}+S_{T2}^{\rm ghost}$. The part made purely from the
gauge superfield~$S_{T2}^{\rm gauge}$ is given by
$$
\eqalign{
   S_{T2}^{\rm gauge}
   &={1\over T(R)}\int d^8z
\cr
   &\quad\times\tr V_Q
   \left[{1\over8}{\cal D}^\alpha\overline D^2{\cal D}_\alpha
         +W_B^\alpha{\cal D}_\alpha
         +{1\over2}({\cal D}^\alpha W_{B\alpha})
         -{\xi\over16}({\cal D}^2\overline D^2
                       +\overline D^2{\cal D}^2)\right]V_Q
\cr
   &=\int d^8z\,V_Q^a
   \left[{1\over8}\widetilde\nabla^\alpha\overline D^2
         \widetilde\nabla_\alpha
         +{1\over2}{\cal W}_B^\alpha\widetilde\nabla_\alpha
         -{\xi\over16}(\widetilde\nabla^2\overline D^2
         +\overline D^2\widetilde\nabla^2)\right]^{ab}V_Q^b,
\cr
}
\eqn\twoxtwenty
$$
where we have shifted to the adjoint representation in the last line.
The field strength in the adjoint representation is defined by
$$
\eqalign{
   &{\cal W}_{B\alpha}\equiv{\cal T}^aW_{B\alpha}^a
   =-{1\over4}\overline D^2(e^{-{\cal V}_B}D_\alpha e^{{\cal V}_B})
   =-{1\over4}[\overline D_{\dot\alpha},
               \{\overline D^{\dot\alpha},\widetilde\nabla_\alpha\}],
\cr
   &\overline{\cal W}_{B\dot\alpha}^\prime
   \equiv{\cal T}^a\overline W_{B\dot\alpha}^{\prime a}
   ={1\over4}e^{-{\cal V}_B}D^2(e^{{\cal V}_B}
   \overline D_{\dot\alpha}e^{-{\cal V}_B})e^{{\cal V}_B}
   ={1\over4}[\widetilde\nabla^\alpha,
              \{\widetilde\nabla_\alpha,\overline D_{\dot\alpha}\}],
\cr
}
\eqn\twoxtwentyone
$$
and in this paper we define the conjugate of the background field
strength as
$$
   \overline W_{B\dot\alpha}^\prime
   \equiv e^{-V_B}\overline W_{B\dot\alpha}e^{V_B}
   ={1\over4}[\nabla^\alpha,\{\nabla_\alpha,
              \overline D_{\dot\alpha}\}]
   \quad{\rm and}\quad
   \overline W_{B\dot\alpha}
   \equiv{1\over4}D^2(e^{V_B}\overline D_{\dot\alpha}e^{-V_B}).
\eqn\twoxtwentytwo
$$
Then by noticing the identity
$$
   \nabla^2\overline D^2+\overline D^2\nabla^2
   -2\nabla^\alpha\overline D^2\nabla_\alpha
   =16\nabla^m\nabla_m+8\overline W_{B\dot\alpha}^\prime
   \overline D^{\dot\alpha}
   +4(\overline D_{\dot\alpha}\overline W_B^{\prime\dot\alpha}),
\eqn\twoxtwentythree
$$
which holds in an arbitrary representation, we find in the
super-Fermi-Feynman gauge~$\xi=1$,
$$
   S_{T2}^{\rm gauge}=
   \int d^8z\,V_Q^a
   \left(-\widetilde\nabla^m\widetilde\nabla_m
   +{1\over2}{\cal W}_B^\alpha\widetilde\nabla_\alpha
   -{1\over2}\overline{\cal W}_{B\dot\alpha}^\prime
   \overline D^{\dot\alpha}\right)^{ab}V_Q^b.
\eqn\twoxtwentyfour
$$

The part of the quadratic action~$S_{T2}$ quadratic in the quantum
chiral superfield is given by
$$
   S_{T2}^{\rm chiral}=
   \int d^8z\,\Phi_Q^\dagger e^{V_B}\Phi_Q
   +\int d^6z\,{1\over2}\Phi_Q^Tm\Phi_Q+{\rm h.c.}
\eqn\twoxtwentysix
$$
We also have mixing terms between the gauge and the chiral
superfields:
$$
\eqalign{
   S_{T2}^{\rm mix}&=
   \int d^8z\,\left(\Phi_B^\dagger e^{V_B}V_Q\Phi_Q
                +\Phi_Q^\dagger e^{V_B}V_Q\Phi_B
                +{1\over2}\Phi_B^\dagger e^{V_B}V_Q^2\Phi_B\right)
\cr
   &\quad+\int d^6z\,g\Phi_B\Phi_Q^2+{\rm h.c.},
\cr
}
\eqn\twoxtwentyseven
$$
where we have included the Yukawa term in~$S_{T2}^{\rm mix}$ for later
convenience.

To second order in quantum fluctuations, the ghost action becomes
$$
\eqalign{
   S_{T2}^{\rm ghost}&=
   {1\over T(R)}\int d^8z\,
   \tr(c^{\prime\dagger}e^{V_B}ce^{-V_B}
   +c^\dagger e^{V_B}c^\prime e^{-V_B}
   +b^\dagger e^{V_B}be^{-V_B})
\cr
   &=\int d^8z\,\left[c^{\prime\dagger a}(e^{{\cal V}_B})^{ab}c^b
          +c^{\dagger a}(e^{{\cal V}_B})^{ab}c^{\prime b}
          +b^{\dagger a}(e^{{\cal V}_B})^{ab}b^b\right].
\cr
}
\eqn\twoxtwentyeight
$$
The similarity of this expression to the action of the chiral
superfield~$\Phi_Q$~\twoxtwentysix\ is obvious. As a consequence,
one-loop quantum effects of ghost fields are simply
obtained by substituting $T^a\rightarrow{\cal T}^a$ and
that of~$\Phi_Q$ by multiplying~$-3$ (with $m=0$ and~$g=0$). Recall
that the ghost fields are Grassmann-odd chiral superfields.

For one-loop level calculations, which we consider in later
sections, it is sufficient to retain only quadratic actions,
$S_{T2}^{\rm gauge}$, $S_{T2}^{\rm chiral}$, $S_{T2}^{\rm ghost}$
and~$S_{T2}^{\rm mix}$. Although we need $S_{T3}$, $S_{T4}$ and so on
for higher loop calculations, further expansion is straightforward.

\chapter{Supersymmetric gauge covariant regularization}
In formulating our regularization scheme, we need a formal propagator
of quantum fields in the presence of the background gauge field~$V_B$.
In the first half of this section, therefore, we explain how to obtain
these propagators. These propagators are defined by inverting the
kinetic operators in $S_{T2}^{\rm gauge}$, $S_{T2}^{\rm chiral}$
and~$S_{T2}^{\rm ghost}$. In other words, we formally diagonalize
these parts of the quadratic action. The remaining parts of the
action, $S_{T2}^{\rm mix}$, $S_{T3}$ and so on, are regarded as the
perturbation. By organizing the perturbative expansion in this way,
we can preserve the background gauge covariance in the regularized
theory, as will be explained in the second half of this section.

\section{Propagators}
It is straightforward to find the propagator of the quantum gauge
superfield~$V_Q$ by formally diagonalizing the quadratic
term~$S_{T2}^{\rm gauge}$~\twoxtwentyfour. The Schwinger-Dyson
equation corresponding to~$S_{T2}^{\rm gauge}$ is
\foot{%
The delta function in the full superspace is denoted as
$\delta(z-z')=\delta(x-x')\delta(\theta-\theta')
\delta(\overline\theta-\overline\theta')$.}
$$
   2\left(-\widetilde\nabla^m\widetilde\nabla_m
   +{1\over2}{\cal W}_B^\alpha\widetilde\nabla_\alpha
   -{1\over2}\overline{\cal W}_{B\dot\alpha}^\prime
   \overline D^{\dot\alpha}\right)^{ab}
   \VEV{T^*V_Q^b(z)V_Q^c(z')}
   =i\delta^{ac}\delta(z-z').
\eqn\threexone
$$
In deriving this expression, we have used the reality
constraint, ${\cal D}^\alpha{\cal W}_{B\alpha}
\equiv\{\widetilde\nabla^\alpha,{\cal W}_{B\alpha}\}
=\overline D_{\dot\alpha}\overline{\cal W}_B^{\prime\dot\alpha}
$~[\WESS]. By formally solving the relation~\threexone, we have
$$
   \VEV{T^*V_Q^a(z)V_Q^b(z')}
   ={i\over2}\left(
   {1\over-\widetilde\nabla^m\widetilde\nabla_m
   +{\cal W}_B^\alpha\widetilde\nabla_\alpha/2
   -\overline{\cal W}_{B\dot\alpha}^\prime
   \overline D^{\dot\alpha}/2}\right)^{ab}\delta(z-z').
\eqn\threextwo
$$
Hereafter, the brackets~$\VEV{\cdots}$ represent an expectation value
in an unconventional interaction picture in which
$S_{T2}^{\rm gauge}$, $S_{T2}^{\rm chiral}$ and~$S_{T2}^{\rm ghost}$
are regarded as the ``un-perturbative part.''

A derivation of the propagator of the quantum chiral
superfield~$\Phi_Q$, on the other hand, is somewhat tricky due to the
chirality constraint. We start with the Schwinger-Dyson equation
derived from~$S_{T2}^{\rm chiral}$~\twoxtwentysix:
$$
\eqalign{
   &-{1\over4}\overline D^2e^{V_B^T}
   \VEV{T^*\Phi_Q^{\dagger T}(z)\Phi_Q^\dagger(z')}
   +m\VEV{T^*\Phi_Q(z)\Phi_Q^\dagger(z')}=0,
\cr
   &-{1\over4}D^2e^{V_B}
   \VEV{T^*\Phi_Q(z)\Phi_Q^\dagger(z')}
   +m^\dagger\VEV{T^*\Phi_Q^{\dagger T}(z)\Phi_Q^\dagger(z')}
   =-{i\over4}D^2\delta(z-z').
\cr
}
\eqn\threexthree
$$
We first multiply $m^\dagger$ from the left of the first relation
in~\threexthree. Then by noting the fact that $m$~is a constant
matrix which satisfies $m^\dagger T^{aT}=-T^am^\dagger$ for the gauge
invariance of the mass term, we find
$$
\eqalign{
   &\left({1\over16}\overline D^2\nabla^2-m^\dagger m\right)
   \VEV{T^*\Phi_Q(z)\Phi_Q^\dagger(z')}
\cr
   &=\left[\nabla^m\nabla_m-{1\over2}W_B^\alpha\nabla_\alpha
   -{1\over4}({\cal D}^\alpha W_{B\alpha})-m^\dagger m\right]
   \VEV{T^*\Phi_Q(z)\Phi_Q^\dagger(z')}
\cr
   &={i\over16}\overline D^2\nabla^2e^{-V_B}\delta(z-z'),
\cr
}
\eqn\threexfour
$$
where the second relation of~\threexthree\ has been used. In going
from the first line to the second line in~\threexfour, we have used
the identity
$$
   \overline D^2\nabla^2+\nabla^2\overline D^2
   -2\overline D_{\dot\alpha}\nabla^2\overline D^{\dot\alpha}
   =16\nabla^m\nabla_m-8W_B^\alpha\nabla_\alpha
   -4({\cal D}^\alpha W_{B\alpha}),
\eqn\threexfive
$$
and the fact that $\Phi_Q$ is chiral
($\overline D_{\dot\alpha}\Phi_Q=0$). Since $\overline D^2$
and~$\nabla^2$ have no inverse, it is generally dangerous to invert
the first line of~\threexfour. (This is obvious by considering the
massless case.) Therefore, we invert the {\it second\/} line
of~\threexfour\ instead to yield
$$
\eqalign{
   &\VEV{T^*\Phi_Q(z)\Phi_Q^\dagger(z')}
\cr
   &={i\over16}\overline D^2
   {1\over\nabla^m\nabla_m-W_B^\alpha\nabla_\alpha/2
   -({\cal D}^\alpha W_{B\alpha})/4-m^\dagger m}\nabla^2e^{-V_B}
   \delta(z-z').
\cr
}
\eqn\threexsix
$$
Here it is important to note that $\overline D^2$ and the inverse
operator in~\threexsix\ commute with each other, as a result of the
identity
$$
\eqalign{
   {1\over16}\overline D^2\nabla^2\overline D^2
   &=\overline D^2
   \left[\nabla^m\nabla_m-{1\over2}W_B^\alpha\nabla_\alpha
   -{1\over4}({\cal D}^\alpha W_{B\alpha})\right]
\cr
   &=\left[\nabla^m\nabla_m-{1\over2}W_B^\alpha\nabla_\alpha
   -{1\over4}({\cal D}^\alpha W_{B\alpha})\right]\overline D^2,
\cr
}
\eqn\threexseven
$$
which follows from~\threexfive. Our derivation of the chiral
propagator~\threexsix, and especially the step from the first line to
the second line in~\threexfour, might appear {\it ad hoc}. However,
our expression~\threexsix\ satisfies the following three criterion:
(1)~It is indeed a solution of the Schwinger-Dyson equation,
\threexthree. (2)~It is manifestly chiral in the $z$~variable and
anti-chiral in the $z'$~variable. (3)~It reduces to the
Grisaru-Ro\v cek-Siegel propagator~[\GRI] when~$V_B=0$. Note that the
Schwinger-Dyson equation~\threexthree\ itself does not ensure the
chirality, and instead the chirality constraint must be supplemented
by hand. In this sense, our expression~\threexsix\ is the unique
solution.

We see below that it is convenient to abbreviate the denominator
of~\threexsix\ as
$$
   {1\over\nabla^m\nabla_m-W_B^\alpha\nabla_\alpha/2
   -({\cal D}^\alpha W_{B\alpha})/4-m^\dagger m}
   \rightarrow
   {1\over\nabla^2\overline D^2/16-m^\dagger m},
\eqn\threexeight
$$
in the sense that
$$
\eqalign{
   &\overline D^2
   {1\over\nabla^m\nabla_m-W_B^\alpha\nabla_\alpha/2
   -({\cal D}^\alpha W_{B\alpha})/4}
   {\nabla^2\overline D^2\over16}
\cr
   &={1\over\nabla^m\nabla_m-W_B^\alpha\nabla_\alpha/2
   -({\cal D}^\alpha W_{B\alpha})/4}
   {\overline D^2\nabla^2\over16}\overline D^2
   =\overline D^2,
\cr
}
\eqn\threexnine
$$
where we have used the identity~\threexfive. The general abbreviation
rule turns to be
$$
   \nabla^m\nabla_m-{1\over2}W_B^\alpha\nabla_\alpha
   -{1\over4}({\cal D}^\alpha W_{B\alpha})
   \leftrightarrow\cases{
   \nabla^2\overline D^2/16&on the right of $\overline D^2$
\cr
   \overline D^2\nabla^2/16&on the left of $\overline D^2$.
\cr}
\eqn\threexten
$$
In what follows, this rule is always understood. For example, a formal
manipulation such as
$$
   \overline D^2{1\over\nabla^2\overline D^2}\nabla^2\overline D^2
   ={1\over\overline D^2\nabla^2}\overline D^2\nabla^2\overline D^2
   =\overline D^2,
\eqn\threexeleven
$$
is legitimate. This kind of manipulation is especially useful in a
calculation of anomalies (see \S5). However, note that
$$
   \overline D^2{1\over\nabla^2\overline D^2}\nabla^2\neq1,
\eqn\threextwelve
$$
because $\overline D^2$ and~$\nabla^2$ themselves have no inverse.
With the abbreviation rule~\threexten, the propagator of the chiral
superfield~\threexsix\ may be written as
$$
   \VEV{T^*\Phi_Q(z)\Phi_Q^\dagger(z')}
   ={i\over16}\overline D^2
   {1\over\nabla^2\overline D^2/16-m^\dagger m}\nabla^2e^{-V_B}
   \delta(z-z').
\eqn\threexthirteen
$$
This form of the propagator is found in the literature (see
Ref.~[\SHIZ] for example). As we have explained, however, this
expression must be used with care. In a similar way, we find the
``Majorana-mass inserted part'',
$$
   \VEV{T^*\Phi_Q(z)\Phi_Q^T(z')}
   ={i\over4}\overline D^2
   {1\over\nabla^2\overline D^2/16-m^\dagger m}m^\dagger
   \delta(z-z').
\eqn\threexfourteen
$$

As was noted below~\twoxtwentyeight, the propagator of ghost fields
which follows from~$S_{T2}^{\rm ghost}$ can be obtained by simply
replacing $T^a\rightarrow{\cal T}^a$ in~\threexthirteen\ (with~$m=0$):
$$
\eqalign{
   &\VEV{T^*c^a(z)c^{\prime\dagger b}(z')}
   =\VEV{T^*c^{\prime a}(z)c^{\dagger b}(z')}
   =\VEV{T^*b^a(z)b^{\dagger b}(z')}
\cr
   &=i\left(\overline D^2{1\over\widetilde\nabla^2\overline D^2}
   \widetilde\nabla^2e^{-{\cal V}_B}\right)^{ab}
   \delta(z-z').
\cr
}
\eqn\threexfifteen
$$
This completes our derivation of propagators which are obtained by
diagonalizing $S_{T2}^{\rm gauge}$, $S_{T2}^{\rm chiral}$
and~$S_{T2}^{\rm ghost}$.

\section{Regularization scheme}
We are now ready to explain our regularization scheme. What we do is
basically an unconventional perturbative expansion in which the parts
of the quadratic action, $S_{T2}^{\rm gauge}$, $S_{T2}^{\rm chiral}$
and~$S_{T2}^{\rm ghost}$, are regarded as the
``unperturbative part.'' The remaining parts of the total action,
$S_{T2}^{\rm mix}$, $S_{T3}$ and so on, are regarded as the
perturbation.\foot{%
When the background scalar superfield~$\Phi_B$ has a vacuum
expectation value, it is necessary to diagonalize the quadratic action
including~$S_{T2}^{\rm mix}$. This generalization should be
straightforward.}
Then the regularization is implemented by substituting for the
propagators, which diagonalize the unperturbative part, with modified
ones.

For example, the propagator of the vector multiplet~\threextwo\ is
replaced by
$$
\eqalign{
   &\VEV{T^*V_Q^a(z)V_Q^b(z')}
\cr
   &\rightarrow
   {i\over2}\Biggl[
   f\left((-\widetilde\nabla^m\widetilde\nabla_m
   +{\cal W}_B^\alpha\widetilde\nabla_\alpha/2
   -\overline{\cal W}_{B\dot\alpha}^\prime
    \overline D^{\dot\alpha}/2)/\Lambda^2\right)
\cr
   &\qquad\qquad\qquad\qquad\quad\times
   {1\over-\widetilde\nabla^m\widetilde\nabla_m
   +{\cal W}_B^\alpha\widetilde\nabla_\alpha/2
   -\overline{\cal W}_{B\dot\alpha}^\prime
   \overline D^{\dot\alpha}/2}
   \Biggr]^{ab}\delta(z-z'),
\cr
}
\eqn\threexsixteen
$$
where $\Lambda$~is the cutoff mass parameter and $f(t)$~is a
regulating factor which decreases sufficiently rapidly:\foot{%
We assume $f(t)$~has a power series expansion in~$t$.}
$$
   f(0)=1,\qquad f(\infty)=f'(\infty)=f''(\infty)=\cdots=0.
\eqn\threexseventeen
$$
Clearly, the prescription~\threexsixteen\ is equivalent to the
proper-time cutoff in the proper-time representation~[\SCH] of the
propagator:
$$
   {i\over2}\left\{
   \int_0^\infty d\tau\,g(\Lambda^2\tau)\exp\left[-\tau
   \left(-\widetilde\nabla^m\widetilde\nabla_m
   +{1\over2}{\cal W}_B^\alpha\widetilde\nabla_\alpha
   -{1\over2}\overline{\cal W}_{B\dot\alpha}^\prime
    \overline D^{\dot\alpha}\right)\right]\right\}^{ab}\delta(z-z'),
\eqn\threexeighteen
$$
where $g(x)$~is the inverse Laplace transformation of~$f(t)/t$. For
example, $g(x)=\theta(x-1)$ for $f(t)=e^{-t}$. By modifying the
propagator in this way, the ultraviolet behavior of Feynman integrals
is tamed, and simultaneously the gauge covariance of the propagator
under the background gauge transformation is preserved. In fact, it
can easily be seen that the propagator, even with the
modification~\threexsixteen, transforms covariantly under the
background gauge transformation~\twoxsix:
$$
   \VEV{T^*V_Q^a(z)V_Q^b(z')}'
   =\left[e^{-i\widetilde\Lambda(z)}\right]^{ac}
   \VEV{T^*V_Q^c(z)V_Q^d(z')}
   \left[e^{i\widetilde\Lambda(z')}\right]^{db},
\eqn\threexnineteen
$$
where $\widetilde\Lambda\equiv{\cal T}^a\Lambda^a$~is the gauge
parameter in the adjoint representation.

Similarly, the propagator of the chiral multiplet~\threexthirteen\ is
replaced by
$$
\eqalign{
   &\VEV{T^*\Phi_Q(z)\Phi_Q^\dagger(z')}
\cr
   &\rightarrow{i\over16}f(-\overline D^2\nabla^2/16\Lambda^2)
   \overline D^2
   {1\over\nabla^2\overline D^2/16-m^\dagger m}\nabla^2e^{-V_B}
   \delta(z-z'),
\cr
}
\eqn\threextwenty
$$
and \threexfourteen\ is replaced by
$$
   \VEV{T^*\Phi_Q(z)\Phi_Q^T(z')}
   \rightarrow{i\over4}f(-\overline D^2\nabla^2/16\Lambda^2)
   \overline D^2
   {1\over\nabla^2\overline D^2/16-m^\dagger m}m^\dagger
   \delta(z-z').
\eqn\threextwentyone
$$
It is easy to see that the propagator~\threextwenty\ transforms as
$$
   \VEV{T^*\Phi_Q(z)\Phi_Q^\dagger(z')}'
   =e^{-i\Lambda(z)}\VEV{T^*\Phi_Q(z)\Phi_Q^\dagger(z')}
   e^{i\Lambda^\dagger(z')},
\eqn\threextwentytwo
$$
under the background gauge transformation~\twoxsix. In \threextwenty\
and~\threextwentyone, we did not include the mass term in the
regulating factor~$f(t)$. Although it would be natural to also include
the mass term in the regulating factor with the proper-time cutoff
prescription, such as~\threexeighteen, it is not required from the
viewpoint of the gauge covariance~\threextwentytwo. Therefore we
omit it from~$f(t)$ for simplicity. The propagators of the ghost
fields~\threexfifteen\ are similarly replaced by
$$
\eqalign{
   &\VEV{T^*c^a(z)c^{\prime\dagger b}(z')}
   =\VEV{T^*c^{\prime a}(z)c^{\dagger b}(z')}
   =\VEV{T^*b^a(z)b^{\dagger b}(z')}
\cr
   &\rightarrow
   i\left[
   f(-\overline D^2\widetilde\nabla^2/16\Lambda^2)
   \overline D^2{1\over\widetilde\nabla^2\overline D^2}
   \widetilde\nabla^2e^{-{\cal V}_B}\right]^{ab}
   \delta(z-z').
\cr
}
\eqn\threextwentythree
$$
Note that the substitutions in \threextwenty, \threextwentyone\
and~\threextwentythree\ preserve the correct chirality.

Now we have to distinguish two cases in evaluating the effective
action or 1PI Green's functions:

(I)~{\it Most\/} 1PI Green's functions are obtained by simply
connecting vertices
in~$S_T-(S_{T2}^{\rm gauge}+S_{T2}^{\rm chiral}+S_{T2}^{\rm ghost})$
by the modified propagators of the quantum fields, \threexsixteen,
\threextwenty, \threextwentyone\ and~\threextwentythree. Recall that
we have formally diagonalized $S_{T2}^{\rm gauge}$,
$S_{T2}^{\rm chiral}$ and~$S_{T2}^{\rm ghost}$ in constructing the
propagators, and thus those parts of the action must be subtracted
from the perturbation. Symbolically, we evaluate the 1PI part of an
expansion of
$$
   \VEV{\exp\left\{i\left[S_T
   -(S_{T2}^{\rm gauge}+S_{T2}^{\rm chiral}+S_{T2}^{\rm ghost})
   \right]\right\}},
\eqn\threextwentyfour
$$
where it is understood that the modified propagators are used. This
defines the first part of the effective action,
${\mit\Gamma}_{\rm I}[V_B,\Phi_B]$.

(II)~However, the above case does not exhaust all. The exception is
1PI diagrams made only from vertices in $S_{T2}^{\rm gauge}$,
$S_{T2}^{\rm chiral}$ and~$S_{T2}^{\rm ghost}$ in the language of the
conventional perturbative expansion. Note that these are necessarily
one-loop diagrams in which all the external lines are the background
gauge superfield. This part of the effective action, which will be
denoted as~${\mit\Gamma}_{\rm II}[V_B]$, corresponds to the logarithm
of the one-loop determinant arising from the Gaussian integration of
$S_{T2}^{\rm gauge}$, $S_{T2}^{\rm chiral}$ and~$S_{T2}^{\rm ghost}$
in our perturbative expansion. However, as is well-known, the
determinant factor cannot be expressed as a one-loop diagram made
from the propagator.\foot{%
A naive ansatz such as~$\VEV{S_{T2}^{\rm chiral}}$, that is, closing
the propagator by another~$S_{T2}^{\rm chiral}$ to form a loop, does
not give the correct combinatorics.}
Therefore we are naturally led to consider a {\it variation\/} of the
effective action instead which can be expressed by a one-loop diagram
with a composite operator inserted:
$$
   {\delta{\mit\Gamma}_{\rm II}[V_B]\over\delta V_B^a(z)}
   \equiv\VEV{J^a(z)},\qquad
   J^a(z)\equiv{\delta\over\delta V_B^a(z)}
   (S_{T2}^{\rm gauge}+S_{T2}^{\rm chiral}+S_{T2}^{\rm ghost}),
\eqn\threextwentyfive
$$
where we have introduced the {\it gauge current superfield\/}~$J^a(z)$
as a variation of the quadratic action.\foot{%
Note that this is a partial gauge current defined with respect to
$S_{T2}^{\rm gauge}$, $S_{T2}^{\rm chiral}$ and~$S_{T2}^{\rm ghost}$,
and not the full gauge current corresponding to the total
action~$S_T$.}
Of course, the modified propagators have to be used
in~\threextwentyfive. We again emphasize that
\threextwentyfive~consists purely of {\it one-loop diagrams in which
all the external lines are the background gauge superfield}.
Therefore, the second part of the effective
action~${\mit\Gamma}_{\rm II}[V_B]$ defined by~\threextwentyfive, if
it exists (see below), is one-loop exact and depends only on the
background vector superfield~$V_B$. The total effective action is
given by a sum:
${\mit\Gamma}[V_B,\Phi_B]={\mit\Gamma}_{\rm I}[V_B,\Phi_B]
+{\mit\Gamma}_{\rm II}[V_B]$.

In addition to the above prescription we have the following:
(III)~A Green's function with a certain composite operator~$O(z)$
inserted is computed by
$$
   \VEV{O(z)\exp\left\{i\left[S_T
   -(S_{T2}^{\rm gauge}+S_{T2}^{\rm chiral}+S_{T2}^{\rm ghost})
   \right]\right\}},
\eqn\threextwentysix
$$
as usual. Again the modified propagators are used.

First of all, it is obvious that with the above prescription, all the
Green's functions are made finite by choosing a sufficiently rapidly
decreasing function~$f(t)$. The high momentum part of the momentum
integration in all the internal loops is suppressed
as~$\sim f(k^mk_m)^n$, where $n$~is the number of internal lines in
a loop. In what follows, furthermore, we show that our scheme
preserves the supersymmetry, the background gauge invariance
(for case~(I)), and the background gauge covariance (for cases (II)
and~(III)).

{\it Supersymmetry\/}:
Since our formulation is expressed entirely by the superfield in an
exactly four dimensional spacetime, the supersymmetry is manifest at
every step. For example, with the above prescription, one can prove
the $N=1$~non-renormalization theorem~[\GRI,\WESS,\WEST] in the form
that the first part of the effective
action~${\mit\Gamma}_{\rm I}[V_B,\Phi_B]$ is a $d^4\theta$ integral
of a product of superfields whose Grassmann coordinates are common,
$\theta$ and~$\overline\theta$. For cases (II) and~(III), the
expectation value of the gauge
current~$\VEV{J^a(x,\theta,\overline\theta)}$, or of the composite
operator~$\VEV{O(x,\theta,\overline\theta)}$, is a product of
superfields whose Grassmann coordinates are common, $\theta$
and~$\overline\theta$. These statements follow from the fact that the
modified propagators \threexsixteen, \threextwenty, \threextwentyone\
and~\threextwentythree\ are expressed by the full superspace delta
function~$\delta(z-z')$ and the fact that the modified propagator of
the chiral multiplet~\threextwenty\ has the $\overline D^2$~factor on
the $z$~variable and the $D^2$~factor on the $z'$~variable. The
latter fact allows one to rewrite interaction vertices in the
superpotential as an integral over the full
superspace~[\GRI,\WESS,\WEST]. Then proof of the
$N=1$~non-renormalization theorem~[\GRI,\WESS,\WEST] can be repeated
even with new unconventional interactions arising from the regulating
factor. This theorem implies, in particular, that
${\mit\Gamma}_{\rm I}[V_B,\Phi_B]$~is supersymmetric invariant, and
the composite operators, $\VEV{J^a(z)}$ or~$\VEV{O(z)}$, are in fact
superfield. Therefore, if ${\mit\Gamma}_{\rm II}[V_B]$ exists as a
functional integration of~$\VEV{J^a(z)}$ with respect to~$V_B(z)$ (see
below), ${\mit\Gamma}_{\rm II}[V_B]$ is supersymmetric invariant too,
and consequently the total effective action is supersymmetric
invariant.

{\it Gauge covariance\/}:
To see how the background gauge invariance and covariance are
preserved with the above prescription, suppose that we perform the
background gauge transformation~\twoxsix\ on $V_B$ and~$\Phi_B$ in a
perturbative expansion of \threextwentyfour, \threextwentyfive\
and~\threextwentysix. Then, at each vertex to which the interaction
term, say~$S_{T3}$, is attached, the background gauge
transformation~\twoxsix\ on $V_B$ and~$\Phi_B$ induces the gauge
transformation on $V_Q$ and~$\Phi_Q$, because the interaction term is
invariant under the background gauge transformation. Note that each
term of the action, $S_{T2}^{\rm mix}$, $S_{T3}$, $S_{T4}$ and so on,
is individually invariant under the background gauge transformation.
Then, those induced transformations on $V_Q$ and~$\Phi_Q$ are canceled
by the transformation at the ends of each propagator line which is
induced by a covariant transformation law such as \threexnineteen\
and~\threextwentytwo. This cancellation does not occur only at a
vertex where a certain operator with a non-trivial gauge
representation is inserted, as \threextwentyfive. In this way, we see
that (at each order of the loop expansion)
${\mit\Gamma}_{\rm I}[V_B,\Phi_B]$ is {\it invariant\/} under the
gauge transformation, and the gauge current~\threextwentyfive\
transforms gauge {\it covariantly\/}:
$$
   \VEV{J^a(z)}'
   ={\partial V_B^b(z)\over\partial V_B'^a(z)}\VEV{J^b(z)}.
\eqn\threextwentyseven
$$
A similar conclusion holds for \threextwentysix: If $O(z)$~has
non-trivial gauge indices, it transforms as expected from the
classical transformation law. In particular, a gauge singlet operator
such as the superconformal current~[\FER] is regularized gauge
{\it invariantly}. The crucial ingredient in the above demonstration
of the gauge covariance is the covariance property of the modified
propagators, \threexnineteen\ and~\threextwentytwo; we have designed
the regulating factors so that this property holds.

For our scheme and the standard proper-time cutoff regularization
(see, for example, the first two references in Ref.~[\NIEL]), there
is an important difference in the treatment of one-loop diagrams. In
our scheme, the second part of the effective
action~${\mit\Gamma}_{\rm II}[V_B]$ is defined through its
{\it variation\/}~$\VEV{J^a(z)}$ in~\threextwentyfive, instead of by a
direct definition, such as the proper-time cutoff in the proper-time
representation of the one-loop {\it determinant}. This apparently
small difference in treatment, however, has a significant consequence.
The point is that one cannot in general define one-loop diagrams of a
chiral fermion in a gauge invariant way, because of the existence of
the gauge anomaly~[\BAR,\NIEL,\PIG]. One thus has to break at some
step the gauge symmetry. How to break this is crucial for the
regularization properties.

As we have seen in~\threextwentyseven, the gauge
current~$\VEV{J^a(z)}$ in our scheme transforms covariantly under the
background gauge transformation. This can be interpreted as implying
that, in a one-loop diagram, the gauge symmetry at all the vertices,
{\it except\/} that of~$J^a(z)$, is preserved. Put a different way, a
possible breaking of the gauge symmetry due to the anomaly is forced
on the $J^a(z)$-vertex. In this way, the gauge symmetry is
``maximally'' preserved in this treatment. On the other hand, in a
``Bose symmetric'' treatment~[\NIEL] such as the proper-time cutoff of
the one-loop determinant or the conventional Pauli-Villars, a breaking
of the gauge symmetry is partitioned equally on every vertex. The
problem with the latter treatment is that it often breaks the gauge
symmetry ``too much.''\foot{%
The important exception is the generalized Pauli-Villars
regularization, which is closely related to our treatment of one-loop
diagrams. (See the last reference in Ref.~[\FUJ] and Okuyama and
Suzuki in Ref.~[\FRO].)}
For example, the conventional Pauli-Villars does not give the
gauge invariant vacuum polarization tensor even for anomaly free
representations ($\tr T^a\{T^b,T^c\}=0$) for which a gauge invariant
regularization would be possible.

Our prescription~\threextwentyfive, on the other hand, respects the
gauge symmetry as much as possible to an extent which does not
contradict with the gauge anomaly. Moreover, when the gauge
representation is free of the gauge anomaly, it is possible to impose
the gauge invariance also on the $J^a(z)$-vertex and, eventually, the
full gauge invariance is restored. In fact, it is possible to show
that the gauge current~$\VEV{J^a(z)}$ has the covariant gauge anomaly:
$$
   -{1\over4}\overline D^2\left\{
   {\cal L}_{V_B/2}
   \cdot\left[\coth({\cal L}_{V_B/2})-1\right]\cdot T^b\right\}^a
   \VEV{J^b(z)}
   \Lambdato-{1\over64\pi^2}\tr T^aW_B^\alpha W_{B\alpha}.
\eqn\threextwentyeight
$$
This is in accord with the gauge covariance
property~\threextwentyseven. We note that this is the {\it exact\/}
expression of the gauge anomaly in our scheme,\foot{%
By introducing an external gauge field which couples to the global
axial current, it might be possible to give a simple proof of the
Adler-Bardeen theorem in a similar way.}
because only one-loop diagrams contribute to~$\VEV{J^a(z)}$
in~\threextwentyfive, and higher-loop diagrams are regularized gauge
invariantly. The left-hand side of~\threextwentyeight\ is a
supersymmetric generalization of the gauge covariant divergence of the
gauge current. This might be interpreted as a gauge non-invariance
of the effective action, $-i\delta{\mit\Gamma}_{\rm II}[V_B^\Lambda]
/\delta\Lambda^a(z)$, in view of the identification~\threextwentyfive.
Equation~\threextwentyeight~shows that the gauge symmetry at the
$J^a(z)$-vertex is restored when~$\tr T^a\{T^b,T^c\}=0$. Therefore,
the full gauge invariance is automatically restored for anomaly-free
cases. This is one of advantages of our scheme. Note that the gauge
multiplet and the ghost multiplet do not contribute
to~\threextwentyeight\
because~$\tr {\cal T}^a\{{\cal T}^b,{\cal T}^c\}=0$.

Our prescription~\threextwentyfive, in place of the manifest gauge
covariance, however, sacrifices the manifest Bose symmetry among gauge
vertices in a one-loop diagram. As a consequence, the second part of
the effective action~${\mit\Gamma}_{\rm II}[V_B]$ whose variation
reproduces the gauge current~$\VEV{J^a(z)}$ may not exist. It is easy
to see that the gauge anomaly must vanish for such a
functional~${\mit\Gamma}_{\rm II}[V_B]$ to exist: When the effective
action exists, the gauge anomaly must satisfy the Wess-Zumino
consistency condition~[\BARD], and thus the gauge anomaly has the
consistent form~[\NIEL,\PIG]. Our anomaly~\threextwentyeight, on the
other hand, has a covariant form. The only possible way out is a zero
covariant anomaly which trivially satisfies the consistency condition.

The reverse is non-trivial: When the gauge anomaly vanishes, is the
Bose symmetry restored, and is it possible to ``integrate'' the
gauge current~$\VEV{J^a(z)}$ to obtain the effective
action~${\mit\Gamma}_{\rm II}[V_B]$ ? We expect the answer is yes, as
is suggested from a restoration of the gauge symmetry at all the
vertices in anomaly-free cases. In fact, our
prescription~\threextwentyfive\ is a natural supersymmetric
generalization of the gauge covariant regularization of
non-supersymmetric chiral gauge theories~[\FUJ,\BAN]. In the
non-supersymmetric theory, when the gauge anomaly vanishes, it
{\it is\/} possible to write down a formal expression of the gauge
invariant effective action whose variation reproduces the covariant
gauge current (in the limit~$\Lambda\rightarrow\infty$)~[\BAN].
Although we postpone to a separate publication~[\HAY] a detailed
analysis of the ``integrability'' of~\threextwentyfive\ (with a
detailed account of~\threextwentyeight), we may expect that the
integrability must be restored, because the supersymmetry should be
irrelevant with regard to this point, and otherwise it would
eventually be impossible to define a gauge invariant effective action
even for anomaly-free cases.

If the integrability is restored in anomaly-free cases, then the
effective action~${\mit\Gamma}_{\rm II}[V_B]$ is supersymmetric and
gauge invariant, because $\VEV{J^a(z)}$~is a gauge covariant
superfield~\threextwentyseven. In particular, a part of the effective
action~${\mit\Gamma}_{\rm II}[V_B]$ corresponding to one-loop diagrams
of the gauge multiplet and the ghost multiplet should always exist,
because the adjoint representation is anomaly-free. The
non-integrability can emerge only in one-loop diagrams of the chiral
multiplet~$\Phi_Q$.\foot{%
In the example of a two-point function of~$V_B$ in the next section,
the non-integrability associated with the gauge anomaly does not
appear, because, as is clear from~\threextwentyeight, the gauge
anomaly is~$O(V_B^2)$ in~$\VEV{J^a(z)}$, which corresponds to a
triangle or a higher-point diagram of~$V_B$.}

In summary, we have observed the following properties of our
regularization scheme. The first part of the effective
action~${\mit\Gamma}_{\rm I}[V_B,\Phi_B]$~\threextwentyfour, which
contains {\it all\/} the 1PI {\it multi-loop\/} diagrams, is always
supersymmetric and gauge invariant. A classically gauge covariant
superfield composite operator, such as the gauge
current~$\VEV{J^a(z)}$ in~\threextwentyfive, remains a gauge covariant
superfield. The breaking of the Bose symmetry and the
non-integrability of~\threextwentyfive\ associated with the gauge
anomaly will be restored for anomaly-free gauge representations
(analysis of this point will be reported elsewhere~[\HAY]). When the
Bose symmetry is restored, we have a supersymmetric invariant, gauge
invariant effective action~${\mit\Gamma}_{\rm II}[V_B]$. This
complication associated with the gauge anomaly, however, occurs only
in one-loop diagrams of~$\Phi_Q$ in which all the external lines are
the background gauge superfield~$V_B$.

\chapter{Illustrative calculations}
In this section we present a somewhat detailed calculation of
one-loop 1PI two-point functions to illustrate how our regularization
scheme works. We believe this demonstration is useful because our
scheme, through the regulating factor, produces new interaction
vertices which do not appear in the conventional super-Feynman rule.
These new vertices ensure the gauge covariance. Since our main
concern in this section is the supersymmetry and the gauge symmetry,
we neglect the effect of the superpotential by setting~$m=g=0$.

To carry out actual calculations, we have to choose a form of the
regulating factor, which should satisfy~\threexseventeen\ and must
be~$O(1/t^\alpha)$ with~$\alpha>1$ for~$t\rightarrow\infty$ to
regulate tadpole diagrams (see below). In this section, we choose as a
simple choice\foot{%
If one can neglect the contribution of tadpoles for some reason, even
use of~$f(t)=1/(t+1)$ may be made. This considerably simplifies the
calculation.}
$$
   f(t)\equiv{1\over(t+1)^2}.
\eqn\fourxone
$$

Our first example is the self-energy part of the chiral multiplet.
Since the external lines of this function are the background chiral
superfield~$\Phi_B$, this case matches category~(I) in the previous
section, \threextwentyfour. At the one-loop level, the self-energy
part is given in configuration space by
$$
\eqalign{
   &\left.\VEV{{\delta^2S_{T2}^{\rm mix}
   \over\delta\Phi_B^\dagger(z)\delta\Phi_B(z')}}\right|_{V_B=0}
   +i\left.
   \VEV{T^*{\delta S_{T2}^{\rm mix}\over\delta\Phi_B^\dagger(z)}
       {\delta S_{T2}^{\rm mix}\over\delta\Phi_B(z')}}\right|_{V_B=0}
\cr
   &={1\over2}T^aT^b\left(-{\overline D^{\prime2}\over4}\right)
              \left[-{D^2\over4}\delta(z-z')\right]
              \left.\VEV{V_Q^a(z')V_Q^b(z')}\right|_{V_B=0}
\cr
   &\quad+i\left(-{D^2\over4}\right)
           \left(-{\overline D^{\prime2}\over4}\right)
   T^a\left.\VEV{T^*\Phi_Q(z)\Phi_Q^\dagger(z')}\right|_{V_B=0}T^b
   \left.\VEV{T^*V_Q^a(z)V_Q^b(z')}\right|_{V_B=0}.
\cr
}
\eqn\fourxtwo
$$
In diagrammatical language, the first term on the right-hand side is
a tadpole. According to the substitution~\threexsixteen, we see that
the tadpole contribution identically vanishes:
$$
\eqalign{
   \left.\VEV{V_Q^a(z')V_Q^b(z')}\right|_{V_B=0}
   &\rightarrow
   {i\over2}\delta^{ab}\lim_{w\rightarrow z'}f(-\dA/\Lambda^2)
   {1\over-\dA}\delta(z'-w)
\cr
   &={i\over2}\delta^{ab}\lim_{w\rightarrow z'}
   {\Lambda^4\over(-\dA+\Lambda^2)^2}{1\over-\dA}\delta(z'-w)
\cr
   &=-{1\over2}\delta^{ab}\int{d^4k\over i(2\pi)^4}\,
   {\Lambda^4\over(k^2+\Lambda^2)^2}{1\over k^2}
   \delta(\theta'-\theta')\delta(\overline\theta'-\overline\theta')
\cr
   &=0,
\cr
}
\eqn\fourxthree
$$
where we have used $\delta(x-y)=\int d^4k\,e^{ik(x-y)}/(2\pi)^4$. As
is well-known, this cancellation is a consequence of the
supersymmetry. However, note that the quadratic divergence is
regularized in~\fourxthree, and a subtlety such as
$\infty\times0$~does not arise.

On the other hand, the second term on the right-hand side
of~\fourxtwo\ leads to
$$
\eqalign{
   &\rightarrow
   iC(R)\left(-{D^2\over4}\right)
        \left(-{\overline D^{\prime2}\over4}\right)
\cr
   &\quad\times\left\{
   \left[{i\over16}f(-\dA/\Lambda^2)
         \overline D^2{1\over\dA}D^2\delta(z-z')\right]
   \left[{i\over2}f(-\dA/\Lambda^2)
         {1\over-\dA}\delta(z-z')\right]\right\}
\cr
   &={i\over2}C(R)\left(-{D^2\over4}\right)
           \left(-{\overline D^{\prime2}\over4}\right)
   \delta(\theta-\theta')\delta(\overline\theta-\overline\theta')
   \left[f(-\dA/\Lambda^2)
         {1\over-\dA}\delta(x-x')\right]^2,
\cr
}
\eqn\fourxfour
$$
where we have used
$$
   \delta(\theta-\theta')\delta(\overline\theta-\overline\theta')
   {\overline D^2D^2\over16}
   \delta(\theta-\theta')\delta(\overline\theta-\overline\theta')
   =\delta(\theta-\theta')\delta(\overline\theta-\overline\theta').
\eqn\fourxfive
$$
Finally, by going to momentum space, we have (recall~\fourxone)
$$
\eqalign{
   &-{1\over2}C(R)\left(-{D^2\over4}\right)
           \left(-{\overline D^{\prime2}\over4}\right)
   \delta(\theta-\theta')\delta(\overline\theta-\overline\theta')
\cr
   &\quad\times\int{d^4p\over(2\pi)^4}\,e^{ip(x-x')}
   \int{d^4k\over i(2\pi)^4}
   {\Lambda^4\over[(k-p)^2+\Lambda^2]^2}
   {\Lambda^4\over(k^2+\Lambda^2)^2}{1\over(k-p)^2}{1\over k^2}
\cr
   &\Lambdato
   -{1\over32\pi^2}C(R)\left(-{D^2\over4}\right)
           \left(-{\overline D^{\prime2}\over4}\right)
   \delta(\theta-\theta')\delta(\overline\theta-\overline\theta')
\cr
   &\qquad\qquad\qquad\qquad\qquad\qquad\qquad\qquad\quad
   \times\int{d^4p\over(2\pi)^4}\,e^{ip(x-x')}
   \left(\ln{\Lambda^2\over p^2}-{5\over6}\right).
\cr
}
\eqn\fourxsix
$$
In terms of the one-loop effective action, \fourxsix\ implies
$$
\eqalign{
   {\mit\Gamma}_{\rm I}^{(1)}[V_B=0,\Phi_B]
   &=-{1\over32\pi^2}C(R)
   \int d^4\theta\int d^4x\,d^4x'\,
   \Phi_B^\dagger(x,\theta,\overline\theta)
   \Phi_B(x',\theta,\overline\theta)
\cr
   &\qquad\qquad\qquad\quad
   \times\int{d^4p\over(2\pi)^4}\,e^{ip(x-x')}
   \left(\ln{\Lambda^2\over p^2}-{5\over6}\right)+\cdots.
\cr
}
\eqn\fourxseven
$$
Apart from the non-universal constant~$-5/6$, which depends on the
precise form of the regulating factor~$f(t)$, \fourxseven~coincides
with the well-known one-loop result~[\WEST]. In fact, since we
{\it know\/} that the first part of the effective
action~${\mit\Gamma}_{\rm I}[V_B,\Phi_B]$ is always gauge invariant,
we can covariantize the local part of the effective action (which is
proportional to~$\ln\Lambda^2$)
as~$\int d^8z\,\Phi_B^\dagger e^{V_B}\Phi_B$, in accord with the
background gauge invariance.

Next, let us consider the vacuum polarization tensor, a one-loop 1PI
two-point function of~$V_B$ ($\Phi_B=0$). This is a typical example
belonging to category~(II), \threextwentyfive. We first study the
contribution of the chiral multiplet. From \twoxtwentysix\
and~\threextwenty, the regularized gauge current is given by
$$
\eqalign{
   &\VEV{J_{\rm chiral}^a(z)}
   \equiv\VEV{{\delta S_{T2}^{\rm chiral}\over\delta V_B^a(z)}}
\cr
   &=\lim_{w\rightarrow z}\tr\VEV{T^*\Phi_Q(z)\Phi_Q^\dagger(w)}
   \left\{T^a+{1\over2}[T^aV_B(w)+V_B(w)T^a]+O(V_B^2)\right\}
\cr
   &\rightarrow
   i\lim_{w\rightarrow z}\tr f(-\overline D^2\nabla^2/16\Lambda^2)
   \overline D^2{1\over\nabla^2\overline D^2}\nabla^2e^{-V_B}
\cr
   &\qquad\qquad\qquad\qquad\qquad\quad\times
   \left[T^a+{1\over2}(T^aV_B+V_BT^a)+O(V_B^2)\right]\delta(z-w).
\cr
}
\eqn\fourxeight
$$
Since we need another $V_B$-line to form the vacuum polarization
tensor, it is sufficient to expand \fourxeight\ in powers of~$V_B$
to~$O(V_B)$. In doing so, we must first recall that the abbreviation
rule~\threexten\ is assumed in~\fourxeight. Then, the expansion
becomes easy by noting the relation~\threexnine. As a result, we have
$$
   \overline D^2{1\over\nabla^2\overline D^2}
   =\overline D^2\left[
   {1\over16\dA}
   -{1\over16\dA}(-V_BD^2\overline D^2+D^2V_B\overline D^2)
    {1\over16\dA}+O(V_B^2)
   \right],
\eqn\fourxnine
$$
and, for the present choice~\fourxone,
$$
\eqalign{
   &f(-\overline D^2\nabla^2/16\Lambda^2)\overline D^2
\cr
   &=\biggl[{\Lambda^4\over(-\dA+\Lambda^2)^2}
   +{\Lambda^4\over(-\dA+\Lambda^2)^2}
   {-\overline D^2V_BD^2+\overline D^2D^2V_B\over16\Lambda^2}
   {\Lambda^2\over-\dA+\Lambda^2}
\cr
   &\qquad\qquad\quad+{\Lambda^2\over-\dA+\Lambda^2}
    {-\overline D^2V_BD^2+\overline D^2D^2V_B\over16\Lambda^2}
    {\Lambda^4\over(-\dA+\Lambda^2)^2}
   +O(V_B^2)\biggr]\overline D^2.
\cr
}
\eqn\fourxten
$$
Substituting \fourxnine\ and~\fourxten\ into~\fourxeight, we have
$$
\eqalign{
   &\VEV{J_{\rm chiral}^a(z)}
   =i\lim_{w\rightarrow z}
   \tr T^a{\Lambda^4\over(-\dA+\Lambda^2)^2}{1\over\dA}
   {\overline D^2D^2\over16}\delta(z-w)
\cr
   &\quad+i\lim_{w\rightarrow z}\tr T^a
    \biggl[
   -{\Lambda^4\over(-\dA+\Lambda^2)^2}{1\over\dA}C
   +{\Lambda^4\over(-\dA+\Lambda^2)^2}C{1\over-\dA+\Lambda^2}
\cr
   &\qquad\qquad\qquad\qquad
   +{1\over-\dA+\Lambda^2}C
   {\Lambda^4\over(-\dA+\Lambda^2)^2}\biggr]
   {\overline D^2D^2\over16}\delta(z-w)+O(V_B^2),
\cr
}
\eqn\fourxeleven
$$
where the combination~$C$ has been defined by
$$
   C\equiv{\overline D^2D^2\over16}V_B{1\over\dA}-V_B.
\eqn\fourxtwelve
$$
Then after some standard spinor algebra, we have in configuration
space,
$$
\eqalign{
   &\VEV{J_{\rm chiral}^a(z)}
   =i\lim_{y\rightarrow x}
   \tr T^a{\Lambda^4\over(-\dA+\Lambda^2)^2}{1\over\dA}
   \delta(x-y)
\cr
   &\quad+i\lim_{y\rightarrow x}\tr T^a
   \biggl[
   -{\Lambda^4\over(-\dA+\Lambda^2)^2}{1\over\dA}C'
   +{\Lambda^4\over(-\dA+\Lambda^2)^2}C'{1\over-\dA+\Lambda^2}
\cr
   &\qquad\qquad\qquad\qquad
   +{1\over-\dA+\Lambda^2}C'
   {\Lambda^4\over(-\dA+\Lambda^2)^2}\biggr]\delta(x-y)+O(V_B^2),
\cr
}
\eqn\fourxthirteen
$$
where the new combination~$C'$ has been defined by
$$
   C'\equiv\left[
   {1\over16}(\overline D^2D^2V_B)
   -{i\over2}(\overline D\overline\sigma^mDV_B)\partial_m
   \right]{1\over\dA}.
\eqn\fourxfourteen
$$

The first line of \fourxeleven\ and~\fourxthirteen\ is~$O(V_B^0)$,
and thus is a tadpole diagram. By going to momentum space, we find a
quadratically divergent Fayet-Iliopoulos $D$-term~[\FAY]:
$$
   \left.\VEV{J_{\rm chiral}^a(z)}\right|_{V_B=0}
   =\tr T^a\Lambda^4\int{d^4k\over i(2\pi)^4}\,
   {1\over(k^2+\Lambda^2)^2}{1\over k^2}
   ={1\over16\pi^2}\tr T^a\Lambda^2.
\eqn\fourxfifteen
$$
Unlike dimensional reduction, in which quadratic divergences always
vanish, our regularization produces this term. However, note that this
term vanishes in conventional models such as the supersymmetric QED,
in which $\tr T^a=0$, so that the gauge-gravitational mixed anomaly
disappears.

The vacuum polarization tensor is given by the $O(V_B)$~term
in~\fourxthirteen. In momentum space, we have
$$
\eqalign{
   &\left.{\delta\VEV{J_{\rm chiral}^a(z)}\over\delta V_B^b(z')}
   \right|_{V_B=0}
   =\tr T^aT^b\int{d^4p\over(2\pi)^4}\,e^{ip(x-x')}
\cr
   &\quad\times\Lambda^4\int{d^4k\over i(2\pi)^4}\,
   \left(
   {1\over16}\overline D^2D^2
   +{1\over2}\overline D\overline\sigma^mDk_m\right)
   \delta(\theta-\theta')
              \delta(\overline\theta-\overline\theta'){1\over k^2}
\cr
   &\qquad\qquad\times\biggl\{
   {1\over[(k+p)^2+\Lambda^2]^2}{1\over(k+p)^2}
\cr
   &\qquad\qquad\qquad
   +{1\over[(k+p)^2+\Lambda^2]^2}{1\over k^2+\Lambda^2}
   +{1\over(k+p)^2+\Lambda^2}{1\over(k^2+\Lambda^2)^2}
   \biggr\}.
\cr
}
\eqn\fourxsixteen
$$
In this expression, the vector derivative~$\partial_m$ in $D_\alpha$
and~$\overline D_{\dot\alpha}$ is understood as~$ip_m$. Note that if
the $\Lambda\rightarrow\infty$~limit is taken in the integrand, only
the first term in the curly bracket survives; this term is what one
would have in the conventional superdiagram calculation. The remaining
terms in~\fourxsixteen\ (the last line) are specific to our scheme,
and these terms ensure the gauge invariance of this two-point
function. The momentum integration of~\fourxsixteen\ is
straightforward in the limit~$\Lambda\rightarrow\nobreak\infty$, and
finally we find
$$
\eqalign{
   &\left.{\delta\VEV{J_{\rm chiral}^a(z)}\over\delta V_B^b(z')}
   \right|_{V_B=0}
\cr
   &\Lambdato{1\over64\pi^2}
   T(R)\delta^{ab}{1\over4}D^\alpha\overline D^2D_\alpha
   \delta(\theta-\theta')
              \delta(\overline\theta-\overline\theta')
   \int{d^4p\over(2\pi)^4}\,e^{ip(x-x')}
   \left(\ln{\Lambda^2\over p^2}+1\right).
\cr
}
\eqn\fourxseventeen
$$
This is the contribution of the chiral multiplet to the vacuum
polarization tensor.

Next let us consider the contribution of the gauge multiplet and the
ghost multiplet. As is well-known~[\GRI], however, the gauge multiplet
cannot contribute to this function at the one-loop level, because the
number of spinor derivatives is not sufficient, as seen from the form
of the propagator~\threextwo\ (it requires at least four~$V_B$). This
is one of advantages of the superfield background field method. This
is also the case even with our modification~\threexsixteen. Therefore
we do not have to evaluate the $V_Q$-loop. No further calculation is
needed also for the ghost's loop, because the contribution of the
ghosts is precisely obtained by $T^a\rightarrow{\cal T}^a$ and by
multiplying the contribution of the chiral multiplet~\fourxseventeen\
by~$-3$, as was noted below~\twoxtwentyeight. Therefore,
from~\fourxseventeen, we have the expression
$$
\eqalign{
   &{\mit\Gamma}_{\rm II}[V_B]
   ={1\over16\pi^2}\Lambda^2\int d^8z\,\tr V_B
\cr
   &+{1\over64\pi^2}[T(R)-3C_2(G)]{1\over2T(R)}
   \int d^4\theta\int d^4x\,d^4x'
\cr
   &\times\tr V_B(x,\theta,\overline\theta)
   \left[{1\over4}D^\alpha\overline D^2D_\alpha
   V_B(x',\theta,\overline\theta)\right]
   \int{d^4p\over(2\pi)^4}\,e^{ip(x-x')}
   \left(\ln{\Lambda^2\over p^2}+1\right)+O(V_B^3),
\cr
}
\eqn\fourxeighteen
$$
completely in terms of the effective action. As discussed in the
previous section, the non-integrability of the gauge
current~$\VEV{J^a(z)}$ associated with the gauge anomaly does not
emerge at this order of the expansion in~$V_B$; actually the
effective action has been obtained to~$O(V_B^2)$ in~\fourxeighteen.
Once the effective action is obtained, we know that it is gauge
invariant, and thus we may covariantize the local term proportional
to~$\ln\Lambda^2$ as~$\int d^6z\,\tr W_B^\alpha W_{B\alpha}$. Of
course, when the matter content has the gauge anomaly
($\tr T^a\{T^b,T^c\}\neq0$) there may exist a finite $O(V_B^2)$~piece
in~$\VEV{J^a(z)}$ which cannot be expressed as a variation of the
effective action. Equation~\fourxeighteen\ again coincides with the
supersymmetric gauge invariant one-loop result~[\WEST], apart from the
non-universal constant~$+1$. The coefficient of~$\ln\Lambda^2$ gives,
as is well-known, the one-loop $\beta$-function of the gauge coupling
constant.

In summary, we have observed that our general discussion on the
supersymmetry and the gauge covariance in the previous section
actually holds in simple but explicit examples. We hope that the
above examples also convinced the reader that our scheme is not too
complicated and that it actually has practical applicability.

\chapter{Super-chiral and superconformal anomalies}
Since our scheme gives a supersymmetric gauge covariant definition of
composite operators, it also provides a simple and reliable way to
compute quantum anomalies. In this section, we present several
examples in the one-loop approximation. Throughout this section, we
assume that the background chiral superfield~$\Phi_B$ and the Yukawa
coupling~$g$ vanish, for simplicity of analysis.

The first example is the super-chiral anomaly~[\KON--\GAT], which is
defined as a breaking of the naive Ward-Takahashi identity:
$$
   -{1\over4}\overline D^2\VEV{\Phi^\dagger e^V\Phi(z)}
   +\VEV{\Phi^Tm\Phi(z)}=0.
\eqn\fivexone
$$
This identity is associated with the chiral symmetry of the massless
action, $\Phi(z)\rightarrow e^{i\alpha}\Phi(z)$, and its explicit
breaking by the mass term. Let us evaluate this anomaly on the basis
of our regularization scheme. We first take in~\fivexone\ the
quadratic terms in the quantum fields (i.e., the one-loop
approximation). Then, as explained in~\threextwentysix, the
regularized super-chiral current is defined by
$$
   \VEV{\Phi_Q^\dagger e^{V_B}\Phi_Q(z)}\equiv
   {i\over16}\lim_{z'\rightarrow z}
   \tr f(-\overline D^2\nabla^2/16\Lambda^2)\overline D^2
   {1\over\nabla^2\overline D^2/16-m^\dagger m}\nabla^2\delta(z-z'),
\eqn\fivextwo
$$
and similarly, from~\threextwentyone,
$$
   \VEV{\Phi_Q^Tm\Phi_Q(z)}\equiv
   {i\over4}\lim_{z'\rightarrow z}
   \tr f(-\overline D^2\nabla^2/16\Lambda^2)\overline D^2
   {1\over\nabla^2\overline D^2/16-m^\dagger m}m^\dagger
   m\delta(z-z').
\eqn\fivexthree
$$
We then directly apply $-\overline D^2/4$ on the composite operator
in~\fivextwo. In so doing, it is important to note that the derivative
acts not only on the $z$~variable but also on the $z'$~variable,
because the equal-point limit is taken prior to the differentiation.
Then, by noting the chirality of~\fivextwo\ with respect to the
$z$~variable and the relation~\threexeleven, we find
$$
\eqalign{
   &-{1\over4}\overline D^2\VEV{\Phi_Q^\dagger e^{V_B}\Phi_Q(z)}
   +\VEV{\Phi_Q^Tm\Phi_Q(z)}
\cr
   &=-{i\over4}\lim_{z'\rightarrow z}
   \tr f(-\overline D^2\nabla^2/16\Lambda^2)
   \overline D^2\delta(z-z')
\cr
   &\Lambdato-{1\over64\pi^2}\tr W_B^\alpha W_{B\alpha}(z),
\cr
}
\eqn\fivexfour
$$
which reproduces the well-known form of the super-chiral
anomaly~[\KON--\GAT]. The details of the calculation
of~\fivexfour~[\KON] are reviewed in the Appendix. We note that the
expression~\fivexfour\ holds even in {\it chiral\/} gauge theories,
and in this sense \fivexfour~may be viewed as a supersymmetric version
of the fermion number anomaly~[\THO]. One might notice that our
calculation of the super-chiral anomaly~\fivexfour\ has resulted in
the Fujikawa method of anomaly evaluation~[\FUJ,\KON]. In fact, the
covariant regularization~[\FUJ] was originally abstracted from the
Fujikawa method.

Since we have defined the regularized composite operator in \fivextwo\
and~\fivexthree, an anomalous supersymmetric commutation relation
associated with the super-chiral anomaly, the Konishi anomaly~[\KONI],
can be derived straightforwardly. First we note in the Wess-Zumino
gauge~[\WESS],
$$
   \Phi^\dagger e^V\Phi
   =A^\dagger A
   +\sqrt{2}\,\overline\theta\overline\psi A+\cdots,
\eqn\fivexfive
$$
and thus classically,
$$
   \sqrt{2}\,\overline\psi^{\dot\alpha}A
   =\overline D^{\dot\alpha}
   \left.(\Phi^\dagger e^V\Phi)\right|_{\theta=\overline\theta=0}.
\eqn\fivexsix
$$
Therefore we may define the supersymmetric transformation of the
composite operator~$\overline\psi^{\dot\alpha}A$ as\foot{%
The normalization of the bracket~$\{,\}$ must be regarded as that of
the (classical) Poisson bracket. For the (anti-){\it commutator}, one
has to multiply the right-hand side of (5.7), (5.23) and~(5.26)
by~$i$.}
$$
\eqalign{
   {1\over2\sqrt{2}}
   \VEV{
   \{\overline Q_{\dot\alpha},\overline\psi_Q^{\dot\alpha}A_Q(x)\}
   }
   &\equiv{1\over4}\overline Q_{\dot\alpha}\overline D^{\dot\alpha}
   \left.\VEV{
   \Phi_Q^\dagger e^{V_B}\Phi_Q(z)}\right|_{\theta=\overline\theta=0}
\cr
   &={1\over4}\overline D^2
   \left.\VEV{
   \Phi_Q^\dagger e^{V_B}\Phi_Q(z)}\right|_{\theta=\overline\theta=0}
\cr
   &=\VEV{A_Q^TmA_Q(x)}
   -{1\over64\pi^2}\tr\lambda_B^\alpha\lambda_{B\alpha}(x).
\cr
}
\eqn\fivexseven
$$
This is the Konishi anomaly~[\KONI] (recall that we set~$\Phi_B=0$).
Our derivation~\fivexseven\ which might appear almost identical to
that in Ref.~[\KON], however, has a conceptual advantage: The point
is that we have first defined the {\it regularized composite
operator}. In this respect, our approach is similar to the original
derivation by Konishi~[\KONI], in which the composite operator is
defined by the gauge invariant point splitting. However, we
regularized the composite operator in terms of the
{\it superfield}. Therefore the supersymmetric transformation of
the regularized composite operator can be performed by one stroke of
the differential operator,
$\overline Q_{\dot\alpha}
=-\partial/\partial\overline\theta^{\dot\alpha}
+i\theta^\alpha\sigma_{\alpha\dot\alpha}^m\partial_m$~[\WESS]. Recall
that we have shown in the previous section that a regularized
composite operator~\threextwentysix\ is in fact a superfield. In this
way, the relation of the Konishi anomaly~\fivexseven\ to the
super-chiral anomaly~\fivexfour\ can be made transparent.

As our next example of the one-loop anomaly, let us consider the
superconformal anomaly~[\FER--\SHI] emerging from the chiral matter
loop. It is a breaking of the Ward-Takahashi relation:
$$
   \overline D^{\dot\alpha}\VEV{R_{\alpha\dot\alpha}(z)}
   -2\VEV{\Phi^Tme^{-V}D_\alpha e^V\Phi(z)}
   +{2\over3}D_\alpha\VEV{\Phi^Tm\Phi(z)}=0.
\eqn\fivexeight
$$
The superconformal current~$R_{\alpha\dot\alpha}$ is defined by
$R_{\alpha\dot\alpha}=R_{\alpha\dot\alpha}^{\rm chiral}
+R_{\alpha\dot\alpha}^{\rm gauge}$, where~[\FER,\SHI]
$$
\eqalign{
   R_{\alpha\dot\alpha}^{\rm chiral}
   &=-\overline D_{\dot\alpha}(\Phi^\dagger e^V)e^{-V}D_\alpha e^V\Phi
   -{1\over3}[D_\alpha,\overline D_{\dot\alpha}]
   (\Phi^\dagger e^V\Phi)
\cr
   &=-\overline D_{\dot\alpha}(\Phi_Q^\dagger e^{V_B})
   \nabla_\alpha\Phi_Q
   -{1\over3}[D_\alpha,\overline D_{\dot\alpha}]
   (\Phi_Q^\dagger e^{V_B}\Phi_Q)
   +\cdots,
\cr
}
\eqn\fivexnine
$$
and~[\FER,\LAN]
$$
   R_{\alpha\dot\alpha}^{\rm gauge}
   =-{2\over T(R)}\tr W_\alpha e^{-V}\overline W_{\dot\alpha}e^V.
\eqn\fivexten
$$
In what follows, we consider only quantum effects of the $\Phi_Q$-loop
and regard the gauge superfield~$V$ as a classical field. In other
words, we shall use the (classical) equation of motion of the gauge
superfield~$V$.

From~\fivexnine, the regularized superconformal current of the chiral
multiplet is defined to the one-loop level by
$$
\eqalign{
   &\VEV{R_{\alpha\dot\alpha}^{\rm chiral}(z)}
\cr
   &\equiv-{i\over16}\lim_{z'\rightarrow z}\tr
   \nabla_\alpha f(-\overline D^2\nabla^2/16\Lambda^2)
   \overline D^2{1\over\nabla^2\overline D^2/16-m^\dagger m}\nabla^2
   \overline D_{\dot\alpha}\delta(z-z')
\cr
   &\quad-{1\over3}{i\over16}[D_\alpha,\overline D_{\dot\alpha}]
   \lim_{z'\rightarrow z}\tr
   f(-\overline D^2\nabla^2/16\Lambda^2)
   \overline D^2{1\over\nabla^2\overline D^2/16-m^\dagger m}\nabla^2
   \delta(z-z').
\cr
}
\eqn\fivexeleven
$$
To deal with this expression, we note the identity
$$
   D_\alpha\lim_{z'\rightarrow z}\tr A(z)\delta(z-z')
   =\lim_{z'\rightarrow z}\tr[\nabla_\alpha,A(z)\}\delta(z-z'),
\eqn\fivextwelve
$$
where $A(z)$~is an arbitrary operator. Then by repeatedly using this
identity and noting the chirality, we find
$$
\eqalign{
   &\VEV{R_{\alpha\dot\alpha}^{\rm chiral}(z)}
\cr
   &={1\over3}{i\over16}\lim_{z'\rightarrow z}\tr
   \biggl[
   -\nabla_\alpha f(-\overline D^2\nabla^2/16\Lambda^2)
   \overline D^2{1\over\nabla^2\overline D^2/16-m^\dagger m}\nabla^2
   \overline D_{\dot\alpha}
\cr
   &\qquad\qquad\qquad\quad
   +f(-\overline D^2\nabla^2/16\Lambda^2)
   \overline D^2{1\over\nabla^2\overline D^2/16-m^\dagger m}\nabla^2
   \overline D_{\dot\alpha}\nabla_\alpha
\cr
   &\qquad\qquad\qquad\quad
   +\overline D_{\dot\alpha}\nabla_\alpha
   f(-\overline D^2\nabla^2/16\Lambda^2)
   \overline D^2{1\over\nabla^2\overline D^2/16-m^\dagger m}\nabla^2
   \biggr]\delta(z-z').
\cr
}
\eqn\fivexthirteen
$$
By applying~$\overline D^{\dot\alpha}$ further, we have
$$
\eqalign{
   &\overline D^{\dot\alpha}
   \VEV{
   R_{\alpha\dot\alpha}^{\rm chiral}(z)
   }
\cr
   &={1\over3}{i\over16}\lim_{z'\rightarrow z}\tr
   \biggl[
   \nabla_\alpha f(-\overline D^2\nabla^2/16\Lambda^2)
   \overline D^2{1\over\nabla^2\overline D^2/16-m^\dagger m}\nabla^2
   \overline D^2
\cr
   &\qquad\qquad\qquad\quad
   -f(-\overline D^2\nabla^2/16\Lambda^2)
   \overline D^2{1\over\nabla^2\overline D^2/16-m^\dagger m}\nabla^2
   \overline D_{\dot\alpha}\nabla_\alpha\overline D^{\dot\alpha}
\cr
   &\qquad\qquad\qquad\quad
   -\overline D^2\nabla_\alpha
   f(-\overline D^2\nabla^2/16\Lambda^2)
   \overline D^2{1\over\nabla^2\overline D^2/16-m^\dagger m}\nabla^2
   \biggr]\delta(z-z').
\cr
}
\eqn\fivexfourteen
$$
Then we use identities
$$
\eqalign{
   &\nabla^2\overline D_{\dot\alpha}\nabla_\alpha
   \overline D^{\dot\alpha}
   =-{1\over2}\nabla^2\overline D^2\nabla_\alpha
   -2\nabla^2 W_{B\alpha},
\cr
   &\overline D^2\nabla_\alpha\overline D^2
   =-4W_{B\alpha}\overline D^2
\cr
}
\eqn\fivexfifteen
$$
to yield
$$
\eqalign{
   &\overline D^{\dot\alpha}
   \VEV{R_{\alpha\dot\alpha}^{\rm chiral}(z)}
   -2\VEV{\Phi_Q^Tm\nabla_\alpha\Phi_Q(z)}
   +{2\over3}D_\alpha\VEV{\Phi_Q^Tm\Phi_Q(z)}
\cr
   &={i\over2}\lim_{z'\rightarrow z}\tr
   \nabla_\alpha f(-\overline D^2\nabla^2/16\Lambda^2)
   \overline D^2\delta(z-z')
\cr
   &\quad-{i\over6}D_\alpha
   \lim_{z'\rightarrow z}\tr
   f(-\overline D^2\nabla^2/16\Lambda^2)
   \overline D^2\delta(z-z')
\cr
   &\quad+2{i\over16}\lim_{z'\rightarrow z}\tr W_{B\alpha}
   f(-\overline D^2\nabla^2/16\Lambda^2)
   \overline D^2{1\over\nabla^2\overline D^2/16-m^\dagger m}\nabla^2
   \delta(z-z'),
\cr
}
\eqn\fivexsixteen
$$
where we have used~\fivextwelve\ again. In this expression, the first
two lines on the right-hand side are regarded as the anomaly, and
the last line can be interpreted as a composite operator. From (A.6)
and~(A.9) in the Appendix, we finally obtain
$$
\eqalign{
   &\overline D^{\dot\alpha}
   \VEV{R_{\alpha\dot\alpha}^{\rm chiral}(z)}
   -2\VEV{\Phi_Q^Tm\nabla_\alpha\Phi_Q(z)}
   +{2\over3}D_\alpha\VEV{\Phi_Q^Tm\Phi_Q(z)}
\cr
   &\Lambdato
   -{1\over8\pi^2}
   \Biggl[\Lambda^2\int_0^\infty dt\,f(t)+{1\over6}\dA\Biggr]
   \tr W_{B\alpha}(z)
   +{1\over192\pi^2}D_\alpha\tr W_B^\beta W_{B\beta}(z)
\cr
   &\quad+2\tr W_{B\alpha}(z)\VEV{\Phi_Q(z)\Phi_Q^\dagger(z)}
   e^{V_B(z)}.
\cr
}
\eqn\fivexseventeen
$$
The presence of the last term is expected, because the classical
equation of motion of the gauge multiplet~$V$ leads to
$$
   \overline D^{\dot\alpha}R_{\alpha\dot\alpha}^{\rm gauge}
   =-2\tr W_\alpha\Phi\Phi^\dagger e^V.
\eqn\fivexeighteen
$$
The right-hand side of this classical expression, when defined to
one-loop accuracy, cancels the last term of~\fivexseventeen.
Therefore, the superconformal anomaly becomes
$$
\eqalign{
   &\overline D^{\dot\alpha}
   \VEV{R_{\alpha\dot\alpha}(z)}
   -2\VEV{\Phi_Q^Tm\nabla_\alpha\Phi_Q(z)}
   +{2\over3}D_\alpha\VEV{\Phi_Q^Tm\Phi_Q(z)}
\cr
   &\Lambdato
   -{1\over8\pi^2}
   \Biggl[\Lambda^2\int_0^\infty dt\,f(t)+{1\over6}\dA\Biggr]
   \tr W_{B\alpha}(z)
   +{1\over192\pi^2}D_\alpha\tr W_B^\beta W_{B\beta}(z),
\cr
}
\eqn\fivexnineteen
$$
when only the one-loop quantum effect of the chiral multiplet is
considered.

We note that although the first term of the superconformal
anomaly~\fivexnineteen\ is perhaps not familiar, it is perfectly
consistent with the anomaly multiplet structure of the superconformal
anomaly~[\FER]; in particular it belongs to the linear anomaly
multiplet~[\CLAR,\SOH]. It is interesting to note that this term
exists only when the gauge-gravitational mixed anomaly exists, i.e.,
when $\tr T^a\neq0$.

Now, as we have done for the Konishi anomaly in~\fivexseven, we may
derive from the superconformal anomaly~\fivexnineteen\ the anomalous
``central extension'' of the $N=\nobreak1$~supersymmetry algebra
which has recently been advocated by Shifman and co-workers~[\DVA]. An
analysis of this problem from the viewpoint of path integrals and
the Bjorken-Johnson-Low prescription is given in Ref.~[\FUJI]. We
first note the definition of the supercharge:
$$
   \overline Q_{\dot\alpha}=\int d^3x\,\overline J_{\dot\alpha}^0,
   \qquad
   \overline J_{\dot\alpha}^m
   =-{1\over2}\overline\sigma^{m\dot\beta\beta}
   \overline J_{\dot\alpha\dot\beta\beta}.
\eqn\fivextwenty
$$
The {\it improved\/}
supercurrent~$\overline J_{\dot\beta\dot\alpha\alpha}$ is related to
the superconformal current~$R_{\alpha\dot\alpha}$ as~[\FER]
$$
   R_{\alpha\dot\alpha}=
   R^{(0)}_{\alpha\dot\alpha}
   -i\overline\theta_{\dot\beta}
   \left(\overline J^{\dot\beta}{}_{\dot\alpha\alpha}
   -{2\over3}\delta_{\dot\alpha}^{\dot\beta}
   \overline J^{\dot\gamma}{}_{\dot\gamma\alpha}\right)+\cdots,
\eqn\fivextwentyone
$$
where the first component~$R^{(0)}_{\alpha\dot\alpha}$ is the
$R$-current. From these relations, we have classically
$$
   \overline J_{\dot\alpha}^0
   =-{i\over2}\overline\sigma^{0\dot\beta\beta}
   \left.(\overline D_{\dot\alpha}R_{\beta\dot\beta}
   -2\varepsilon_{\dot\alpha\dot\beta}\overline D^{\dot\gamma}
   R_{\beta\dot\gamma})\right|_{\theta=\overline\theta=0}.
\eqn\fivextwentytwo
$$
Therefore we may define the supersymmetric transformation of the
supercurrent operator as
$$
\eqalign{
   \VEV{
   \{\overline Q_{\dot\alpha},\overline J_{\dot\beta}^0(x)\}
   }
   &\equiv
   -{i\over2}\overline\sigma^{0\dot\gamma\gamma}
   \overline Q_{\dot\alpha}
   \left.\left[\overline D_{\dot\beta}\VEV{R_{\gamma\dot\gamma}(z)}
   -2\varepsilon_{\dot\beta\dot\gamma}\overline D^{\dot\delta}
   \VEV{
   R_{\gamma\dot\delta}(z)
   }\right]\right|_{\theta=\overline\theta=0}
\cr
   &={i\over2}
   \overline\sigma^{0\dot\gamma\gamma}
   (\varepsilon_{\dot\alpha\dot\beta}\overline D_{\dot\gamma}
   +2\varepsilon_{\dot\beta\dot\gamma}\overline D_{\dot\alpha})
   \left.\overline D^{\dot\delta}
   \VEV{R_{\gamma\dot\delta}(z)}\right|_{\theta=\overline\theta=0}
\cr
   &=-{1\over48\pi^2}
   \overline\sigma^{0m}_{\dot\alpha\dot\beta}\partial_m
   \tr\lambda_B^\alpha\lambda_{B\alpha}(x),
\cr
}
\eqn\fivextwentythree
$$
where we have used the superconformal anomaly~\fivexnineteen.\foot{%
We have set~$m=0$ for simplicity.}
The ``central extension'' of the $N=1$~supersymmetry algebra~[\DVA] is
obtained by integrating~\fivextwentythree\ over the spatial
coordinate~$x$. (Recall that we have taken into account only the
chiral matter loop in~\fivexnineteen.) In deriving the second line
from the first line in~\fivextwentythree, we have used the identity
(see the last reference of~[\DVA])
$$
   \left.\overline D_{\dot\alpha}\overline D_{\dot\beta}
   X_{\gamma\dot\gamma}\right|_{\theta=\overline\theta=0}
   =-\varepsilon_{\dot\alpha\dot\beta}
   \left.\overline D_{\dot\gamma}\overline D^{\dot\delta}
   X_{\gamma\dot\delta}\right|_{\theta=\overline\theta=0},
\eqn\fivextwentyfour
$$
which may be confirmed component by component. From the above
derivation, it is clear that use of the improved
supercurrent~\fivextwentyone\ and the presence of the superconformal
anomaly~\fivexnineteen\ are crucial to have the anomalous term.
Although our derivation in~\fivextwentythree\ is apparently almost
identical to that in Ref.~[\DVA], it is conceptually quite
transparent, as we have explained for the Konishi anomaly~\fivexseven.

In a similar way, we may study another anomalous supersymmetric
transformation law, the commutator between the supercharge and the
$R$-current (or charge)~[\ITO]. The $R$-current is defined from the
superconformal current~\fivextwentyone\ by
$$
\eqalign{
   R^{(0)m}
   &=-{1\over2}\overline\sigma^{m\dot\alpha\alpha}
   R^{(0)}_{\alpha\dot\alpha}
\cr
   &=-{1\over2}\overline\sigma^{m\dot\alpha\alpha}
   \left.R_{\alpha\dot\alpha}\right|_{\theta=\overline\theta=0}.
\cr
}
\eqn\fivextwentyfive
$$
Then the supersymmetric transformation of the $R$-current is related
to the superconformal anomaly as
$$
\eqalign{
   \VEV{[\overline Q_{\dot\alpha},R^{(0)0}(x)]}
   &\equiv-{1\over2}\overline\sigma^{0\dot\beta\beta}
   \left.\overline Q_{\dot\alpha}\VEV{R_{\beta\dot\beta}(z)}
   \right|_{\theta=\overline\theta=0}
\cr
   &=-i\VEV{\overline J_{\dot\alpha}^0(x)}
   -\overline\sigma^{0\dot\beta\beta}
   \varepsilon_{\dot\alpha\dot\beta}
   \left.\overline D^{\dot\gamma}\VEV{R_{\beta\dot\gamma}(z)}
   \right|_{\theta=\overline\theta=0}
\cr
   &=-i\VEV{\overline J_{\dot\alpha}^0(x)}
   +{i\over8\pi^2}
   \Biggl[\Lambda^2\int_0^\infty dt\,f(t)+{1\over6}\dA\Biggr]
   \tr\lambda_B^\alpha(x)\sigma_{\alpha\dot\alpha}^0
\cr
   &\qquad\qquad
   -{i\over48\pi^2}\tr\lambda_B^\alpha(x)
   \left[{1\over2}\sigma_{\alpha\dot\alpha}^0D_B(x)
   +\sigma_{\alpha\dot\alpha}^mv_B^{+0}{}_m(x)\right],
\cr
}
\eqn\fivextwentysix
$$
where we have used the relation~\fivextwentytwo\ and the
superconformal anomaly~\fivexnineteen.\foot{%
We have again set~$m=0$ for simplicity.}
In the last line, $v_{mn}^+$~is the self-dual part of the field
strength:
$$
   v_{mn}^+\equiv
   {1\over2}\left(v_{mn}+{i\over2}\varepsilon_{mnkl}v^{kl}\right).
\eqn\fivextwentyseven
$$
In Ref.~[\ITO], an anomalous term in the commutator~\fivextwentysix\
is analyzed in the supersymmetric pure Yang-Mills theory with use of
the gauge invariant point splitting regularization. Although the
structure of our result~\fivextwentysix\ is not quite the same as that
of Ref.~[\ITO] (the anomalous term is given by the dual of the field
strength instead of the self-dual part), the discrepancy seems to
originate from the definition of the regularized $R$-current (see
also Ref.~[\FUJI]).

Similar calculations are possible also for the superconformal anomaly
emerging from the gauge multiplet loop~[\CLA,\GRISA,\SHIF]. We will
report on this part of analysis elsewhere~[\HAY].

\chapter{Conclusion}
In this paper, we have formulated a manifestly supersymmetric
gauge covariant regularization of supersymmetric chiral gauge
theories. As we have shown, this scheme has many desired features
which are, we believe, not shared by other single regularization
schemes proposed to this time.\foot{%
Strictly speaking, the integrability of~\threextwentyfive\ in
anomaly-free cases remains to be proven to definitely conclude the
existence of the effective action in such cases (although we are
almost sure of this point from experience in non-supersymmetric
theories~[\BAN]). This analysis regarding the integrability will be
reported elsewhere~[\HAY].} Certainly, we must admit that we did
not mention the unitarity of the $S$-matrix which, with the BRST (or
quantum gauge) symmetry, is not manifest in our regularization
scheme. This point is concerned with the issue of how the effective
action in the background field method and the $S$-matrix are
related~[\DEW,\ABB] {\it at the regularized level}. We are at present
unable to make a general statement on this point. However, it is
obvious that the background gauge invariance of the effective action
(or 1PI diagrams) is certainly a {\it necessary condition\/} for the
unitarity of the $S$-matrix, because otherwise the longitudinal mode
does not decouple from the $S$-matrix. We emphasize that our scheme,
to our knowledge, is the first attempt which fulfills this necessary
condition for supersymmetric {\it chiral\/} gauge theories.

We have presented several illustrative applications, but only in the
one-loop approximation. The situation for multi-loop diagrams in our
scheme, on the other hand, is rather simple, because the supersymmetry
and the gauge invariance of the effective action are always ensured.
Although this property is similar to that of the supersymmetric
higher covariant derivative regularization~[\WES], our calculation
rule is reasonably simple compared to the higher derivative
regularization. Therefore it is of great interest to see how actual
higher loop calculations proceed in our scheme. We hope to come back
this problem in the near future.

We thank Dr.~H.~Igarashi for numerous discussions and for his
collaboration in the early stages of this work. H.~S. is grateful to
Professor K.~Fujikawa and Kazumi Okuyama for helpful correspondence
on a related topic. The work of H.~S. is supported in part by the
Ministry of Education Grant-in-Aid Scientific Research,
Nos.~09740187, 09226203 and~08640348.

\appendix

In this appendix, we explain the evaluation of the anomalous factors,
$$
   \lim_{z'\rightarrow z}\tr\left\{{1\atop\nabla_\alpha}\right\}
   f(-\overline D^2\nabla^2/16\Lambda^2)
   \overline D^2\delta(z-z').
\eqn\axone
$$
Our calculation is basically a generalization of that in Ref.~[\KON].
However, the actual calculation is somewhat quicker, thanks to our
manifestly gauge covariant treatment. First, let us recall that the
abbreviation rule~\threexten\ is assumed in~\axone. Next, we note the
momentum representation of the delta function and
$$
\eqalign{
   &e^{-ikx}\overline D_{\dot\alpha}e^{ikx}
   =\overline D_{\dot\alpha}
   +\theta^\alpha\sigma_{\alpha\dot\alpha}^mk_m,
\cr
   &e^{-ikx}\nabla_me^{ikx}
   =\nabla_m+ik_m,
\cr
   &e^{-ikx}\nabla_\alpha e^{ikx}
   =\nabla_\alpha-\sigma_{\alpha\dot\alpha}^m
   \overline\theta^{\dot\alpha}k_m,
\cr
}
\eqn\axtwo
$$
and thus \axone\ can be expressed as
$$
\eqalign{
   &\tr\Lambda^4\int{d^4k\over(2\pi)^4}\,
   \left\{{1\atop
   \nabla_\alpha-\sigma_{\alpha\dot\alpha}^m
   \overline\theta^{\dot\alpha}\Lambda k_m}\right\}
\cr
   &\quad\times f\biggl(
   k^nk_n-2ik^n\nabla_n/\Lambda
   -W_B\sigma^n\overline\theta k_n/(2\Lambda)
\cr
   &\qquad\qquad\quad
   -\nabla^n\nabla_n/\Lambda^2
   +W_B^\beta\nabla_\beta/(2\Lambda^2)
   +({\cal D}^\beta W_{B\beta})/(4\Lambda^2)\biggr)
\cr
   &\qquad\quad\times
   (\overline D_{\dot\gamma}
   +\theta^\gamma\sigma_{\gamma\dot\gamma}^k\Lambda k_k)
   (\overline D^{\dot\gamma}
   +\theta^\delta\sigma_\delta^{l\dot\gamma}\Lambda k_l)
   \left.
   \delta(\theta-\theta')\delta(\overline\theta-\overline\theta')
   \right|_{\theta=\theta',\overline\theta=\overline\theta'},
\cr
}
\eqn\axthree
$$
where we have rescaled the momentum variable
as~$k_m\rightarrow\Lambda k_m$. Then we expand~$f(\cdots)$ in inverse
powers of~$\Lambda$. At this step, it should be noted that at least
two spinor derivatives have to act on the delta function to survive:
$$
   \overline D^2\left.\delta(\overline\theta-\overline\theta')
   \right|_{\overline\theta=\overline\theta'}=-4,
   \qquad
   \nabla_\alpha\nabla_\beta\left.\delta(\theta-\theta')
   \right|_{\theta=\theta'}=-2\varepsilon_{\alpha\beta}.
\eqn\axfour
$$
But also note that three are too many:
$$
   \nabla_\alpha\nabla_\beta\nabla_\gamma=0.
\eqn\axfive
$$
Then it is easy to see that the first case of~\axthree\ gives~[\KON]
$$
\eqalign{
   &\lim_{z'\rightarrow z}\tr f(-\overline D^2\nabla^2/16\Lambda^2)
   \overline D^2\delta(z-z')
\cr
   &\Lambdato
   -\int{d^4k\over(2\pi)^4}\,f''(k^mk_m)\tr W_B^\alpha W_{B\alpha}
\cr
   &=-{i\over16\pi^2}\tr W_B^\alpha W_{B\alpha},
\cr
}
\eqn\axsix
$$
where we have used the value of the momentum integration
$$
   \int{d^4k\over(2\pi)^4}\,
   \left\{
   \matrix{f'(k^mk_m)\cr f''(k^mk_m)\cr f'''(k^mk_m)k^nk^k\cr}
   \right\}
   ={i\over16\pi^2}
   \left\{
   \matrix{-\int_0^\infty dt\,f(t)\cr
           1\cr -\eta^{nk}/2\cr}
   \right\},
\eqn\axseven
$$
which is useful also for the second case in~\axthree.

On the other hand, the necessary expansion is somewhat lengthy in the
second case of~\axthree, and we do not reproduce it here. It might be
useful, however, to note the following: The straightforward expansion
of~\axthree\ contains terms explicitly proportional to
$\overline\theta^2$ and~$\overline\theta_{\dot\alpha}$ which, at first
glance, do not seem to vanish. If these terms survive, the right-hand
side of~\axthree\ cannot be a superfield, because the lowest component
($\theta=\overline\theta=0$~term) of these terms vanishes (and thus
the higher components must vanish if it is a superfield~[\WESS]). In
fact, it can be confirmed that the terms explicitly proportional to
$\overline\theta^2$ and~$\overline\theta_{\dot\alpha}$ vanish by
noting relations such as
$$
   \tr W_{B\alpha}W_B^\beta W_{B\beta}=0,
   \qquad
   \nabla_m W_{B\alpha}W_{B\beta}-\nabla_m W_{B\beta}W_{B\alpha}
   =\varepsilon_{\alpha\beta}\nabla_m W_B^\gamma W_{B\gamma},
\eqn\axeight
$$
where the first relation follows from the cyclic property of the
trace.

Finally, after some rearrangements, we obtain
$$
\eqalign{
  &\lim_{z'\rightarrow z}\tr\nabla_\alpha
   f(-\overline D^2\nabla^2/16\Lambda^2)
   \overline D^2\delta(z-z')
\cr
   &\Lambdato
   {i\over4\pi^2}
   \Biggl[\Lambda^2\int_0^\infty dt\,f(t)+{1\over6}\dA\Biggr]
   \tr W_{B\alpha}
   -{i\over32\pi^2}D_\alpha\tr W_B^\beta W_{B\beta}.
\cr
}
\eqn\axnine
$$

\refout
\bigskip
\titlestyle{\bf Note added}
By using the technique presented in the Appendix, it is easy to obtain
the {\it divergent part\/} of the gauge current~\threextwentyfive\
for an {\it arbitrary} $f(t)$:
$$
\eqalign{
   &\Lambda{d\over d\Lambda}\VEV{J^a(z)}
\cr
   &\Lambdato{\delta\over\delta V^a(z)}\left[
   {\Lambda^2\over8\pi^2}\int_0^1dt\,f(t)\int d^8z\,\tr V_B
   +{T(R)-3C_2(G)\over64\pi^2}\int d^6z\,\tr W_B^\alpha W_{B\alpha}
   \right].
\cr
}
$$
This shows that the divergent part is always (i.e., whether or not the
gauge anomaly is present) integrable, and the one-loop
$\beta$-function is actually independent of $f(t)$.

\bye